\documentclass[fleqn,usenatbib]{mnras}

\usepackage{newtxtext,newtxmath}

\usepackage[T1]{fontenc}

\DeclareRobustCommand{\VAN}[3]{#2}
\let\VANthebibliography\thebibliography
\def\thebibliography{\DeclareRobustCommand{\VAN}[3]{##3}\VANthebibliography}

\usepackage{graphicx}	
\usepackage{amsmath}	
\usepackage{booktabs}
\usepackage{tabularx}
\usepackage{adjustbox}
\usepackage{upgreek}
\usepackage{anyfontsize}
\usepackage{subcaption}

\newcommand{\unif}[2]{\ensuremath{\mathcal{U} (#1,#2)}}

\newcommand{\trades}[0]{\textsc{TRADES}}
\newcommand{\pyde}[0]{\textsc{PyDE}}
\newcommand{\emcee}[0]{\textsc{emcee}}

\title[]{A 34.6-day transiting sub-Neptune in the TOI-1422 planetary system}

\author[L. Naponiello et al]{
L. Naponiello$^{1}$\thanks{E-mail: luca.naponiello@inaf.it}, %
P. Leonardi$^{2,3}$, 
M. Damasso$^{1}$, 
M.-L. Steinmeyer$^{4}$, 
M. Stalport$^{5,6}$, 
C. Dorn$^{4}$, 
\newauthor
A.~S.~Bonomo$^{1}$, 
L. Mancini$^{7,1,8}$, 
A. Sozzetti$^{1}$, 
S. Benatti$^{9}$, 
S. Colombo$^{9}$, 
and R. Cosentino$^{10}$ 
\\ \\
$^{1}$INAF -- Osservatorio Astrofisico di Torino, Via Osservatorio 20, 10025 Pino Torinese, Italy\\
$^{2}$Dipartimento di Fisica, Università di Trento, Via Sommarive 14, I-38123 Povo, Italy\\
$^{3}$Dipartimento di Fisica e Astronomia, Università degli Studi di Padova, Vicolo dell’Osservatorio 3, I-35122 Padova, Italy\\
$^{4}$Institute for Particle Physics and Astrophysics, ETH Zürich, Otto-Stern-Weg 5, 8093 Zürich, Switzerland\\
$^{5}$Space sciences, Technologies and Astrophysics Research (STAR)
Institute, Université de Liège, Allée du 6 Août 19C, 4000 Liège,
Belgium\\
$^{6}$Astrobiology Research Unit, Université de Liège, Allée du 6 Août
19C, 4000 Liège, Belgium\\
$^{7}$Department of Physics, University of Rome ``Tor Vergata'', Via della Ricerca Scientifica 1, 00133 Rome, Italy\\
$^{8}$Max Planck Institute for Astronomy, K\"{o}nigstuhl 17, 69117 Heidelberg, Germany \\
$^{9}$INAF -- Osservatorio Astronomico di Palermo, Piazza del Parlamento 1, 90134, Palermo, Italy\\
$^{10}$Fundaci\'{o}n Galileo Galilei - INAF, Rambla Jos\'{e} Ana Fernandez P\'{e}rez 7, 38712 Bre\~{n}a Baja, TF, Spain\\
}

\date{Accepted 2025 November 13. Received 2025 November 13; in original form 2025 September 04.}

\pubyear{\the\year{}}

\begin{document}
\label{firstpage}
\pagerange{\pageref{firstpage}--\pageref{lastpage}}
\maketitle

\begin{abstract}
TOI-1422 is a G2\,V star ($V = 10.6$ mag) known to host a warm Neptune-sized planet, TOI-1422\,b, with a mass and radius of about $9\,M_{\oplus}$ and $4\,R_{\oplus}$,  on a circular orbit with a period of $12.997$ days. An outer planetary candidate in this system had previously been suggested on the basis of a residual signal in the radial velocity (RV) data with a tentative period of $\sim$29 days, along with a possible single transit-like event, although it was not clear at the time whether the two signals belonged to the same companion. In this work, we confirm the presence of a second transiting planet, TOI-1422\,c, a sub-Neptune ($R=2.61\pm0.14\,R_{\oplus}$) that orbits with a longer period of 34.563 days. This confirmation is based on the detection of three TESS transits, two from newly available sectors, combined with new and archival RV measurements. The sub-Neptune ($\rho_{\rm c}=4.3^{+1.3}_{-1.0}$ g\,cm$^{-3}$) is more massive than the inner Neptune ($\rho_{\rm b}=0.93^{+0.21}_{-0.20}$ g\,cm$^{-3}$), having a mass of $M_{\rm c}=14\pm3\,M_{\oplus}$, making TOI-1422 a rare anti-ordered system. 
Furthermore, we detect transit timing variations (TTVs) on the inner planet, with amplitudes of up to 5 hours, suggesting ongoing dynamical interactions. A dynamical analysis that combined TTVs and RVs indicates that planet c alone is unlikely to account for the full TTV amplitude observed on TOI-1422\,b.
We investigated whether an additional, as yet undetected companion could account for the observed signal, exploring a range of plausible orbital configurations and finding that a low-mass planet located between the two known orbits may be responsible.
\end{abstract}




\section{Introduction}

The architectural arrangement of planets within a single system provides one of the most stringent tests of planet-formation theory. While the properties of an individual planet constrain its nature, the relative ordering of masses and radii among sibling planets acts as a direct fossil record of their collective origin and evolution \citep{Winn2015}. From this comparative approach, two general features have emerged for mature systems: ($i$) an intra-system uniformity known as ``peas in a pod'', i.e. planets in the same system tend to be similar in size and regularly spaced \citep{ChatterjeeTan2014,Millholland2017,Weiss2018}, and ($ii$) a predictable (non-linear) mass-radius relation where larger planets are also more massive \citep{Fortney2007,LopezFortney2014,ChenKipping2017}. Systems that violate these general patterns offer invaluable clues about the non-canonical, often extreme, processes that shape planetary architectures. A notable exception to the aforementioned trends occurs in the small subset of multiplanet systems where the inner planet is substantially larger in size but less massive than its companion on a wider orbit (see, e.g., the list compiled by \citealt{Howe2025}). 

A particular case is that of the two warm mini-Neptunes in the TOI-815 system \citep{Psaridi2024}, where the inner planet, TOI-815\,b, has both a larger radius and a lower mass than the outer planet TOI-815\,c.
This configuration presents a challenge to quiescent formation models, which struggle to explain how a smaller, denser, and more massive body could form or end up in an orbit exterior to a larger, less massive ``puffy'' sibling. Explaining this configuration points toward a more dynamic formation history than is typically assumed. The observed architecture could be the result of gravitational scattering events that re-ordered the system, transporting planets from different birth locations \citep{Izidoro2015,Bitsch2019}. Alternatively, the outer planet may have suffered a catastrophic giant impact late in its evolution, which would have stripped away a primordial gas envelope, leaving behind a smaller, denser and now more massive core \citep{Liu2015,Bonomo2019,Psaridi2024}. Both scenarios imply a departure from the smooth processes that are believed to shape most of planetary systems.

In this paper, we report the discovery of a second transiting planet orbiting TOI-1422, a quiet G2\,V star ($V = 10.6$\,mag) 155 pc away, previously known to host a warm Neptune-sized planet ($P_{\rm orb}=12.997$\,days; \citealt{Naponiello2022}). The new planet, a sub-Neptune, orbits externally to the first and is smaller in size, yet our analysis reveals that it is the more massive of the two. Furthermore, significant transit timing variations (TTVs) detected for the inner planet, unrelated to the new one, suggest that the system is not yet fully characterized and may host an additional unseen perturber. The system, therefore, represents a compelling natural laboratory for studying complex evolutionary histories of exoplanets. 

The paper is structured as follows. In Sect.~\ref{sec:observations}, we describe the new photometric and spectroscopic observations. In Sect.~\ref{sec:analysis}, we present the system parameters derived from our global analysis that confirm this unusual hierarchy. We explore the composition of both planets in Sect.~\ref{sec:formation}, discuss the properties of the system in Section \ref{sec:discussion} and draw conclusions in Section \ref{sec:conclusions}.

\section{Observations and data reduction}\label{sec:observations}
\subsection{TESS photometry}\label{sec:TESS}
TIC\,333473672 (TOI-1422) has been observed by TESS in late 2019 (Sectors 16 and 17), in October 2022 (Sector 57) and, recently, in October 2024 (Sector 84). We retrieved the two-minute cadence photometry of all four sectors from the Mikulski Archive for Space Telescopes (MAST), as reduced by the Science Processing Operations Center (SPOC; \citealt{Jenkins2016}) pipeline, developed at the NASA Ames Research Center. In particular, we used the Presearch Data Conditioning Simple Aperture Photometry (PDC-SAP; \citealt{Stumpe2012, Stumpe2014}, \citealt{Smith2012}) light curve, which is already corrected for dilution and systematics. We found no appreciable short- or long-term modulation in both the PDC-SAP and SAP light curves, in accordance with the notion that the star is rather old and quiet \citep{Naponiello2022}.

\subsection{Radial velocities}\label{sec:harpn}

The original RV data set was obtained from 112 spectra taken with the High Accuracy Radial velocity Planet Searcher for the Northern hemisphere (HARPS-N; \citealt{Cosentino2012}) instrument at the Telescopio Nazionale Galileo (TNG) in La Palma (Spain), within the GAPS Neptune program \citep{Naponiello2022}. On top of that, we included 9 new spectra recently taken with the same instrument, as part of the ``Ariel Masses Survey (ArMS)'' large program (Filomeno et al. 2025 in review, Di Maio et al. in prep.), dedicated to the precise mass measurement of transiting planets with radii spanning the radius valley potentially valuable for atmospheric characterization with the ESA Mission Ariel \citep{Tinetti_2018}. The new observations were carried out to refine the ephemerides of TOI-1422\,b, which is considered a promising Ariel target with a transmission spectroscopy metric of $\sim$100, and to investigate the origin of the linear trend reported in \citet{Naponiello2022}.

The RVs and activity indices (Table\,\ref{tab:RVs}) were extracted using version 3.0.1 of the HARPS-N Data Reduction Software ({\tt DRS}, \citealt{Dumusque2021}). In total, after removing one outlier\footnote{This RV point has the largest uncertainty ($\sim9$\,m\,s$^{-1}$) and corresponds to the spectrum with the lowest signal-to-noise ratio of our data set, which also yields anomalous values in all the activity indices.}, we used 120 RVs with an average uncertainty of $3.0$\,m\,s$^{-1}$, a weighted root-mean-square (wrms) of $\sim5$\,m\,s$^{-1}$, and a signal-to-noise ratio (SNR) of \,$\approx$\,45, measured at a reference wavelength of 5500 {\AA}. Recently, 34 RV measurements of TOI-1422, with an average uncertainty of $2.1$\,m\,s$^{-1}$ (but with a large wrms of $\sim7$ m/s), were released \citep{Polanski2024}. However, we did not include the RVs obtained by the High Resolution Echelle Spectrometer (HIRES; \citealt{Vogt1994}) in this work due to their limited number and sparse temporal sampling (one observation every $\sim28$ days).

The updated HARPS-N RV data set now reveals a quadratic trend extending beyond the observational timespan ($\sim1700$ days), although it could be stellar in origin as it closely resembles the shape of two activity indicators (see Figs.\,\ref{fig:long_trend}, \ref{fig:crosscorr}), despite the star being rather quiet, with a median chromospheric index of $\log{R^{\prime}_{\rm HK}}=-4.942\pm0.005$. While we refer to the discovery paper \citep{Naponiello2022} for a careful analysis of the activity diagnostics (the new dataset only includes 9 new RVs taken a few years later), we report in Fig.\,\ref{fig:activity} the GLS periodograms of the $\log{R^{\prime}_{\rm HK}}$ and H$\alpha$ indices, which show a long-term modulation similar to that of the RVs.

We treated this trend employing a simple quadratic term in the global analysis that follows (see Sect.\,\ref{sec:globalfit}), rather than using a Gaussian Process (GP; \citealt{Rajpaul2015}). The long-term variation is well defined and could also arise from a combination of stellar activity and a genuine long-period planetary signal, so we considered it safer and more transparent to model it as a deterministic trend, especially since GPs can be overly flexible and partially absorb low-amplitude planetary signals (e.g. \citealt{Blunt2023}).

\section{Analysis}\label{sec:analysis}
\subsection{GLS periodogram}

\begin{figure}
\centering
\includegraphics[width=0.48\textwidth]{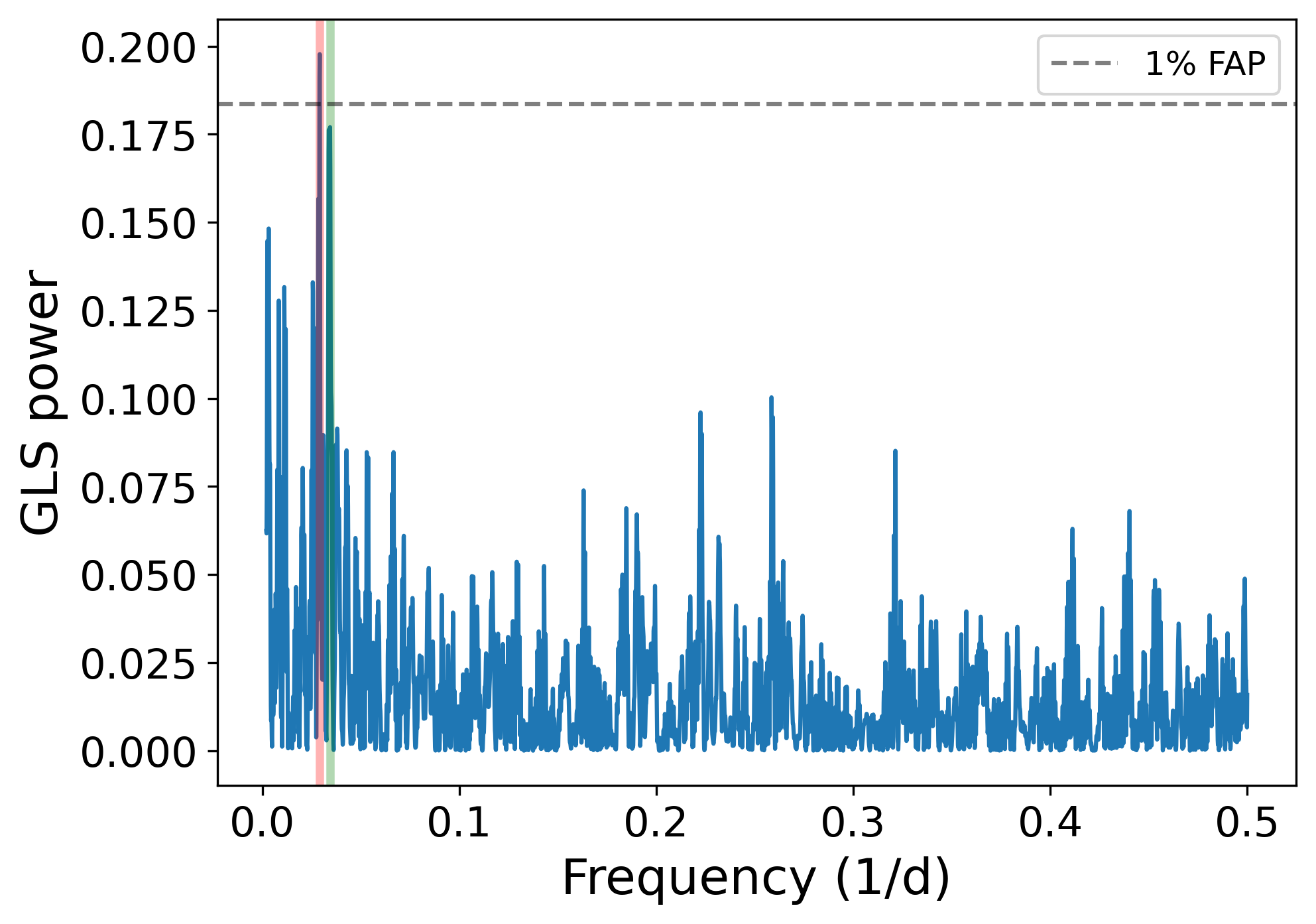}
\caption{GLS periodogram of the RV residuals from the 1-planet model, once the quadratic trend has been removed. The red and green vertical lines signal, respectively, the periodicity of $34.6$ and $29.0$\,days, while the horizontal dashed line marks the 1\% FAP.}\label{fig:periodogram}
\end{figure}

We computed the generalized Lomb-Scargle (GLS) periodogram for the HARPS-N residual RVs, after removing both the long quadratic term and the TOI-1422\,b signal. We found that the highest power does not correspond to $\sim29$\,d, the signal announced by \citealt{Naponiello2022}, but rather to  $\sim34.6$\,d, one of the two aliases indicated in Fig.\,D2 of the same paper, with a low false alarm probability (FAP) of 0.2\% (Fig.\,\ref{fig:periodogram}). However, these two signals are easily recognized as aliases of each other when combined with the second highest frequency of the window function ($1/190$\,d) between 1.5 and 1000 days. Therefore, this information alone is not sufficient to determine whether one of these signals is of planetary origin, nor to establish which one is genuine.

\subsection{Photometric periodograms}\label{sec:periodograms}

A single transit-like signal, which meets the minimum diagnostic criteria\footnote{In particular, the event was transit-shaped with resolved ingress and egress, showing no correlation with the X–Y pixel position or background flux, and exhibiting a consistent depth across different photometric apertures. The signal was also recovered independently with multiple photometric pipelines, including those employing alternative neighbour-subtraction techniques \citep{Naponiello2022}.}, was first reported in the discovery paper near $2\,458\,756.35\,{\rm BJD_{TDB}}$, yet the authors argued that it was not compatible with the 29\,d RV signal as ``\textit{the lack of other transits in the TESS light curve suggests an orbital period between 17-22, or longer than $\sim35$ days}''. With the addition of two TESS sectors and the 34.6\,d signal gaining ground over the 29\,d one in the new RV dataset, we searched the entire TOI-1422 light curve for new transits using the Cambridge Exoplanet Transit Recovery Algorithm (CETRA; \citealt{Smith2025}), a fast and sensitive transit detection algorithm optimised for graphic processing units (GPUs). In particular, after masking TOI-1422\,b transits, we have searched for transits with a minimum depth of 50 parts per million (ppm). 

The algorithm identified a peak with a period of 34.56\,d, corresponding to a transit duration of 6.9 hours, a transit depth of $683\pm48$\,ppm, and a $\mathrm{SNR}\sim14.3$, with the centre of the first transit being pinpointed at $2\,458\,756.367\,{\rm BJD_{TDB}}$, precisely matching the single transit previously reported. We repeated the analysis using the more commonly used box least squares (BLS; \citealt{Kovacs2002}) and transit least squares (TLS; \citealt{Hippke2019}) periodograms, finding the same periodicity, although with lower significance. In total, three transits have been observed for this new candidate, one for each TESS sector, with the exclusion of Sector 17. For TOI-1422\,b, the number of transits amounts to 7 at the time of writing, compared to 4 of the discovery paper. The host star is not scheduled for observation by TESS in the upcoming sectors.

\subsection{Global fit}\label{sec:globalfit}

\begin{figure}
\centering
\includegraphics[width=0.48\textwidth]{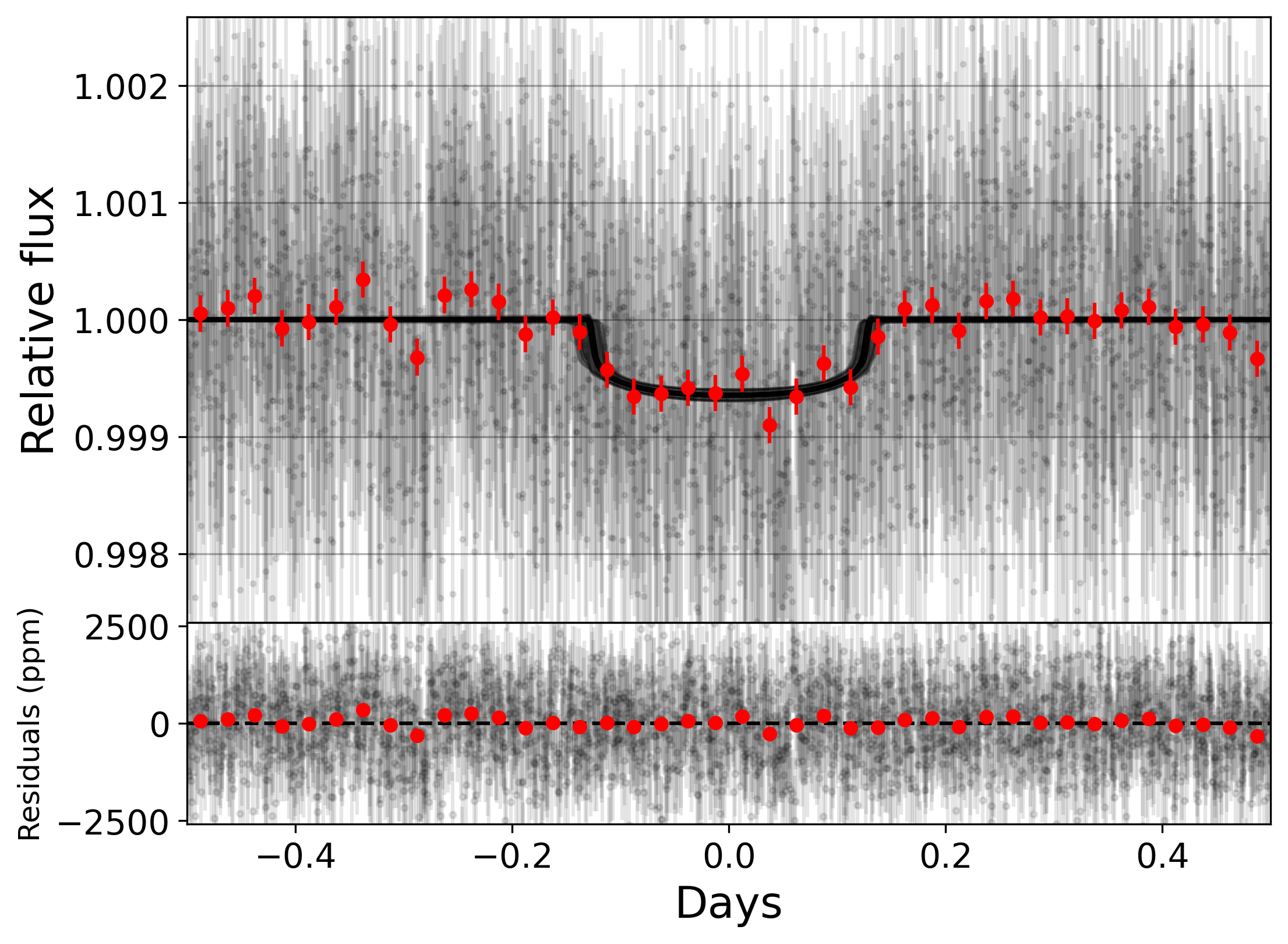}
\includegraphics[width=0.48\textwidth]{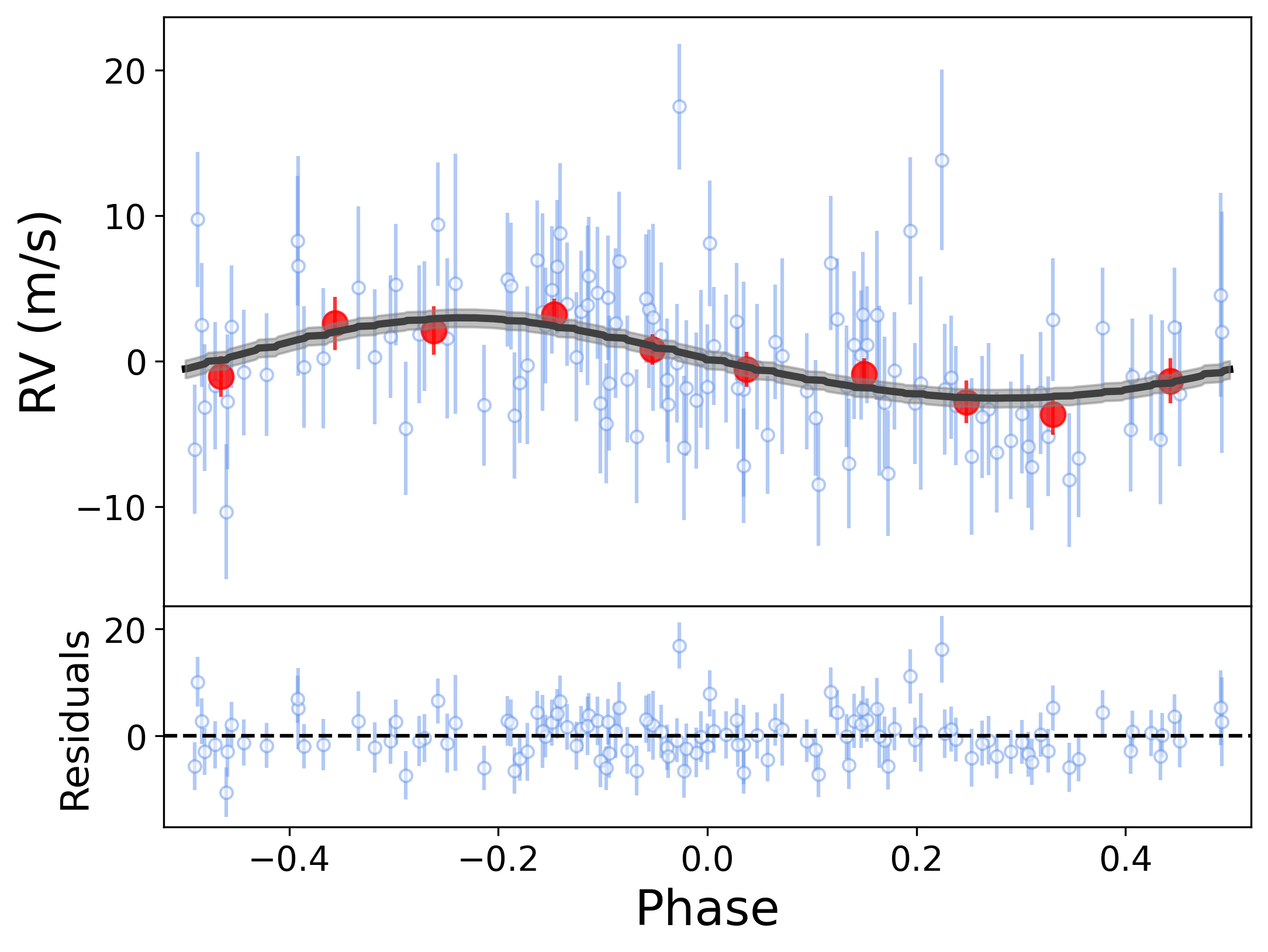}
\caption{Phased TESS transits (top panel) and RV signal (bottom panel) of TOI-1422\,c, along with the best fitted models, in black, and their residuals below each panel. The red circles represent the average of $\sim31$ minutes and $\sim83$ hours, respectively, while the gray areas represent the $1\sigma$ deviation from each model.}
\label{fig:toi1422c_phased}
\end{figure}

For homogeneity with the previous work, we have performed a joint transit and RV analysis employing \texttt{juliet} \citep{Espinoza2019} and using the same approach described by \citet{Naponiello2022}. Mainly, we have tried different models with a combination of 1 or 2 planets and fixed to zero or free eccentricities. We found that the 2-planet model with circular orbits is more significant than the 2-planet model with free eccentricities ($\Delta\ln{\mathcal{Z}}^{e=0}_{e\geq0}=6.9$), which, at the same time, is much more significant than any 1-planet model ($\Delta\ln{\mathcal{Z}}^{\mathrm{2p}}_{\mathrm{1p}}>10$; where $\mathcal{Z}$ is the Bayesian evidence of each fit). Therefore, we consider TOI-1422\,c as a new confirmed planet with $P_{\rm orb}=34.5633\pm0.0002$\,d, $M_{\rm c}=14\pm3\,M_{\oplus}$, and $R_{\rm c}=2.61\pm0.14\,R_{\oplus}$. The phased RV signal and the transits of TOI-1422\,c have been plotted in Fig.\,\ref{fig:toi1422c_phased}, while in Fig.\,\ref{fig:fullRV} we show the entire RV data set. We also refined the parameters of TOI-1422\,b (see Table\,\ref{tab:2p}), which are mostly in agreement with the previous estimates.

Finally, the GLS periodogram of the adopted model’s residuals (Fig. \ref{fig:periodogram_2p}) reveals a signal at about 387\,d (FAP $\lesssim1\%$). However, a subsequent 3-planet model failed to properly account for this signal, and we therefore do not consider it significant. We stress that this signal's origin is unclear: while it is close to the possible 1-year Earth alias, it also corresponds to a potential periodicity found in our dynamical analysis (Sect. \ref{sec:dynamical}).

\begin{table*}
\centering
\caption{Priors and posteriors for the global analysis of TOI-1422. Priors are listed for fitted parameters, while derived parameters are calculated from the posteriors of the fitted model.}
\label{tab:2p}
\renewcommand{\arraystretch}{1.2}
\begin{tabular}{l l c c}
    \toprule
    \textbf{} & \textbf{Prior} & \textbf{TOI-1422\,b} & \textbf{TOI-1422\,c} \\
    \midrule
    \multicolumn{4}{l}{\textbf{Stellar parameters}} \\[2pt]
    Stellar density, $\rho_{\star}$ (kg\,m$^{-3}$)\dotfill & $\mathcal{N}(1300,\:100)^{(\mathparagraph)}$ & \multicolumn{2}{c}{$1277\pm88$} \\
    Systemic velocity, $\gamma$ (m\,s$^{-1}$)\dotfill & $\mathcal{U}(-25957,\:-25937)$ & \multicolumn{2}{c}{$-25948^{+7}_{-6}$} \\[2pt]
    \midrule
    \multicolumn{4}{l}{\textbf{Transit and orbital parameters}} \\[2pt]
    Orbital period, $P_{\rm orb}$ (days)\dotfill & $\mathcal{N}(12.99, 0.01)$ (b) & $12.99894\pm0.00003$ & $34.5633\pm0.0002$ \\
    & $\mathcal{U}(15, 100)$ (c) & & \\
    Time of transit, $T_{\rm 0}$ (BJD$-2458700$)\dotfill & $\mathcal{N}(45.92, 0.01)$ (b) & $45.9405^{+0.0023}_{-0.0021}$ & $56.3525^{+0.0062}_{-0.0057}$ \\
    & $\mathcal{N}(56.351, 0.01)$ (c) & & \\
    RV semi-amplitude, $K$ (m\,s$^{-1}$)\dotfill & $\mathcal{U}(0, 10)$ & $2.64^{+0.54}_{-0.53}$ & $2.80\pm0.59$ \\
    $r_1^{(\S)}$\dotfill & $\mathcal{U}(0,1)$ & $0.0345\pm0.0009$ & $0.023\pm0.001$ \\
    $r_2^{(\S)}$\dotfill & $\mathcal{U}(0,1)$ & $0.19\pm0.12$ & $0.17^{+0.14}_{-0.12}$\\
    Eccentricity, $e$\dotfill & Fixed at 0\,$^{(*)}$ & $<0.20^{(\ddagger)}$ & $<0.20^{(\ddagger)}$ \\
    Argument of periastron, $\omega$ ($\deg$)\dotfill & Fixed at 90\,$^{(*)}$ & unconstrained & unconstrained \\
    Parametrized limb-darkening coeff., $q_1$\dotfill & $\mathcal{N}(0.31, 0.30)^{(**)}$ & \multicolumn{2}{c}{$0.40^{+0.21}_{-0.17}$} \\
    Parametrized limb-darkening coeff., $q_2$\dotfill & $\mathcal{N}(0.25, 0.10)^{(**)}$ & \multicolumn{2}{c}{$0.25\pm0.09$} \\[2pt]
    \midrule
    \multicolumn{4}{l}{\textbf{Derived parameters}} \\[2pt]
    Total transit duration, $T_{\rm 14}$ (hr)\dotfill & Derived & $4.46\pm0.12$ & $6.12\pm0.18$ \\
    Orbital inclination, $i$ ($\deg$)\dotfill & Derived & $89.53^{+0.30}_{-0.31}$ & $89.77^{+0.15}_{-0.19}$ \\
    Impact parameter, $b$\dotfill & Derived & $0.19\pm0.12$ & $0.17^{+0.14}_{-0.12}$ \\
    Scaled semi-major axis, $a/R_{\star}$\dotfill & Derived & $22.51^{+0.50}_{-0.53}$ & $43.20^{+0.97}_{-1.02}$ \\
    Planet radius, $R_{\rm p}$ ($R_{\oplus}$)\dotfill & Derived & $3.83\pm0.11$ & $2.61\pm0.14$ \\
    Planet mass, $M_{\rm p}$ ($M_{\oplus}$)\dotfill & Derived & $9.5^{+2.0}_{-1.9}$ & $14\pm3$ \\
    Planet density, $\rho_{\rm p}$ (g\,cm$^{-3}$)\dotfill & Derived & $0.93^{+0.21}_{-0.20}$ & $4.3^{+1.3}_{-1.0}$ \\
    Equilibrium temperature, $T_{\rm eq}^{(\dagger)}$ (K)\dotfill & Derived & $870\pm14$ & $628\pm10$ \\
    Surface gravity, $\log{g_{p}}$ (cgs)\dotfill & Derived & $6.37^{+1.39}_{-1.34}$ & $20.12^{+5.09}_{-4.56}$ \\
    Semi-major axis, $a$ (AU)\dotfill & Derived & $0.107\pm0.003$ & $0.205\pm0.005$ \\
    Limb-darkening coeff., $u_1$\dotfill & Derived & \multicolumn{2}{c}{$0.31^{+0.15}_{-0.13}$} \\
    Limb-darkening coeff., $u_2$\dotfill & Derived & \multicolumn{2}{c}{$0.30^{+0.15}_{-0.13}$} \\[2pt]
    \midrule
    \multicolumn{4}{l}{\textbf{Instrumental parameters}} \\[2pt]
    HARPS-N RV jitter, $\sigma_{\textsf{HARPS-N}}$ (m\,s$^{-1}$)\dotfill & $\mathcal{U}(0, 10)$ & \multicolumn{2}{c}{$3.5\pm0.4$} \\
    RV trend coeff., $A^{(***)}$ (m\,s$^{-1}$\,d$^{-1}$)\dotfill & $\mathcal{U}(0, 0.05)$ & \multicolumn{2}{c}{$0.024\pm0.005$} \\
    RV trend coeff., $B^{(***)}$ (m\,s$^{-1}$)\dotfill & $\mathcal{U}(-200, 0)$ & \multicolumn{2}{c}{$-13\pm7$} \\
    RV trend coeff., $C^{(***)}$ (m\,s$^{-1}$\,d$^{-2}$)\dotfill & $\mathcal{U}(-0.1, 0.1)$ & \multicolumn{2}{c}{$(8\pm2)\cdot10^{-6}$} \\
    TESS white noise, $\sigma_{\textsf{TESS}}$ (ppm)\dotfill & $\mathcal{L}(10^{-2}, 10^2)$ & \multicolumn{2}{c}{$0.5^{+9.0}_{-0.5}$} \\
    TESS GP amplitude, $\sigma_{\textsf{GP,\,TESS}}$ (ppm)\dotfill & $\mathcal{L}(10^{-6}, 10)$ & \multicolumn{2}{c}{$0.13\pm0.02$} \\
    TESS GP timescale, $\rho_{\textsf{GP,\,TESS}}$ (days)\dotfill & $\mathcal{L}(10^{-2}, 10)$ & \multicolumn{2}{c}{$0.78^{+0.15}_{-0.20}$} \\
    \bottomrule
\end{tabular}
\begin{flushleft}
$^{(\mathparagraph)}$ We adopt the stellar density prior derived from the stellar characterization presented in \citealt{Naponiello2022}. \\
$^{(\S)}$ ($r_1$, $r_2$) is the parametrization described in \protect\citet{Espinoza2018} for $R_p/R_{\star}$ and the impact parameter $b$. \\
$^{(\dagger)}$ This is the equilibrium temperature for a zero Bond albedo and uniform heat redistribution to the night side. \\
$^{(*)}$ In the case of non-null eccentricity, the priors were set as follows: $(\sqrt{e}\,\sin\omega, \sqrt{e}\,\cos\omega)$ in $\mathcal{U}(-1.0,\:1.0)$. \\ 
$^{(**)}$ The limb-darkening priors come from the theoretical values of \citealt{Claret2017}, following \citealt{Naponiello2022}. \\
$^{ (\ddagger)}$ The eccentricity upper limit is constrained at the confidence level of 1$\sigma$. \\
$^{(***)}$ $A$, $B$ and $C$ are coefficients of the long-term trend which is added to the Keplerian model and is of the form: $B+A(t-t_i)+C(t-t_i)^2$.
\end{flushleft}
\end{table*}

\subsection{Transit Timing Variations}
In an isolated Keplerian system, planetary transits occur with clockwork regularity. Any significant deviation from this periodicity (TTVs), indicates the presence of additional dynamical or physical processes at play.
Such variations are most often driven by gravitational perturbations from other planets \citep{Agol2005, Holman2005} but can also result from secular effects acting on close-in giant planets, including tidal orbital decay \citep{rasio_1996, patra_2017, Leonardi_2024} or apsidal precession \citep{Ragozzine_Wolf_2009}.

For stars that are bright enough, it is also possible to obtain high-precision RV measurements that, combined with TTVs, allow alleviation of their individual biases and posterior degeneracies \citep{Nespral_2017, Malavolta_2017, Petigura_2018, Nascimbeni_2023, Borsato_2024, Korth_2024}.

\begin{figure*}
\centering
\includegraphics[width=1\textwidth]{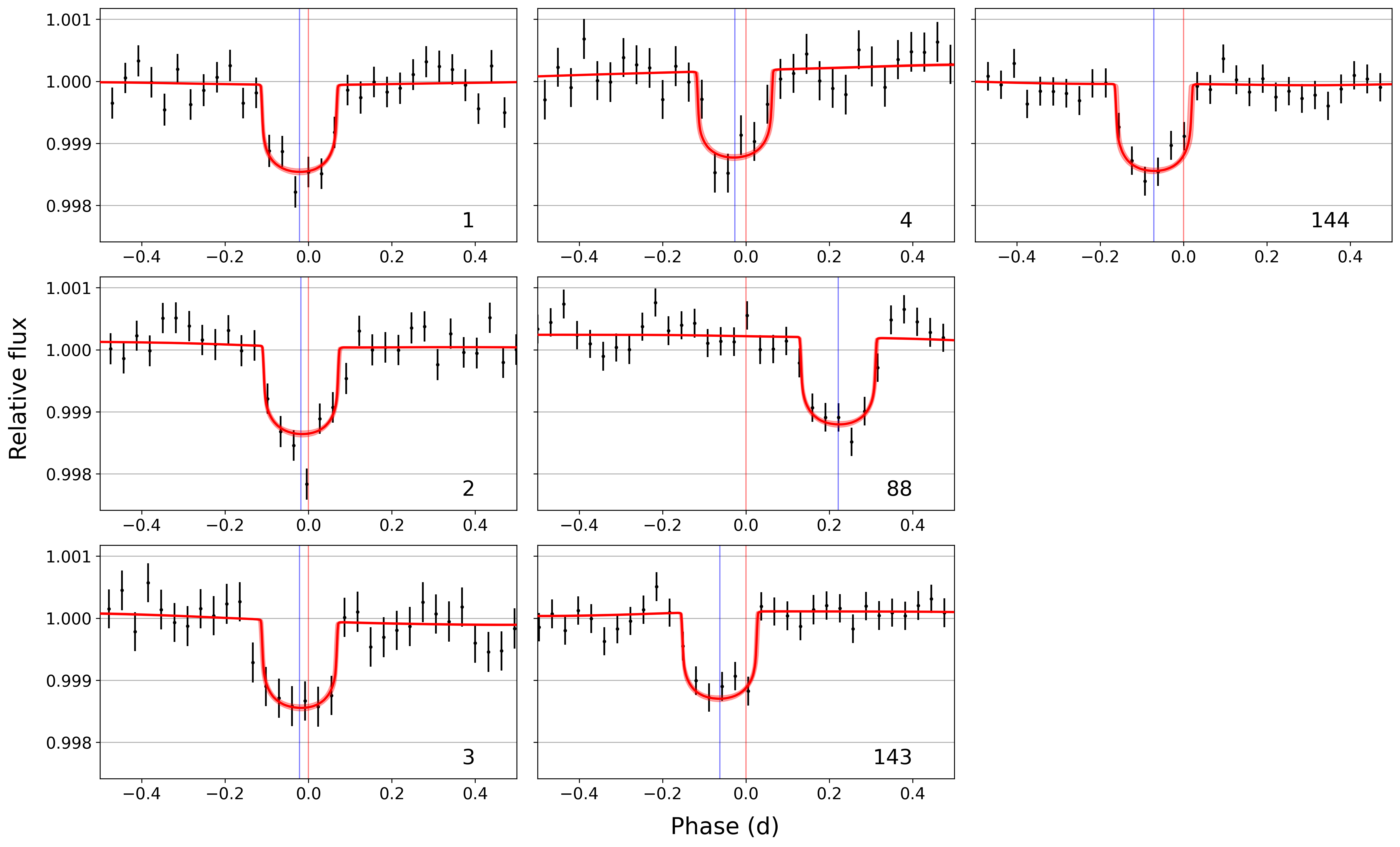}
\caption{TTVs of TOI-1422\,b. The epochs of the transits are labeled on the bottom right of each panel. The red and blue vertical lines indicate, respectively, the calculated and observed center times of the transits.}\label{fig:TTV_planetb}
\end{figure*}

\begin{figure*}
\centering
\includegraphics[width=1\textwidth]{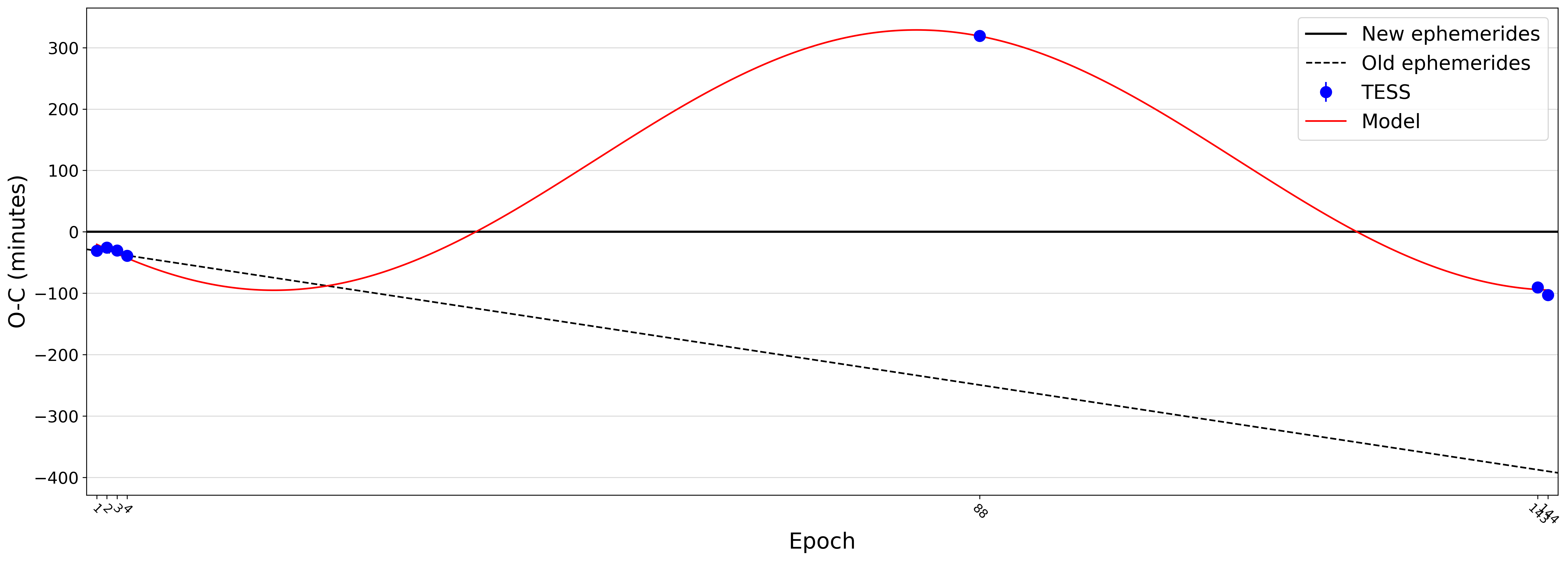}
\caption{Observed minus calculated (O-C) center times of transits for TOI-1422\,b. The ephemerides published in the discovery paper are represented by the dashed line, while the black solid line represents the ephemerides as evaluated in this manuscript. The red line is a simple sinusoidal fit to the O-C times. The error bars of the O-C times are smaller than the represented blue circles.}\label{fig:OC}
\end{figure*}

No TTVs were detected in the discovery paper of TOI-1422\,b, as only four transits of the Netpune-sized planet had been observed consecutively in TESS Sectors 16 and 17. In contrast, our updated analysis, benefiting from an extended time baseline, reveals a significant deviation especially in the timing of the fifth detected transit of TOI-1422\,b (epoch 88, observed in Sector 57). By fitting each mid-transit time independently (Fig.\,\ref{fig:TTV_planetb}), we find that this transit occurs more than 8 hours later than predicted by a linear ephemeris (Fig.\,\ref{fig:OC}). A similar analysis was performed for TOI-1422\,c. However, with only three observed transits, no significant timing deviations were detected (Fig.~\ref{fig:TTV_planetc}).
Given the separation between the planets (period ratio $P_{\rm c}/P_{\rm b} \approx 2.65$) TOI-1422\,b and TOI-1422\,c are not in or near a strong low-order mean motion resonance (MMR), which could help to explain the observed TTVs amplitude. Such a large deviation, therefore, strongly suggests gravitational perturbations from an additional, as yet undetected, planetary companion.

\subsection{Dynamical modeling: TTVs + RVs}\label{sec:dynamical}
To investigate which dynamical interactions can be responsible for the observed timing variations, we employed the N-body integrator included in \texttt{TRADES}\footnote{\url{https://github.com/lucaborsato/trades}} \citep{Borsato2014, Borsato2019, Borsato2021}, which simultaneously fits the transit times and RV measurements (see Table~\ref{tab:RVs}), while integrating the system parameters.

We tested four different dynamical configurations for the system: (1) a two-planet model including only the known transiting planets; (2) a three-planet model with an additional inner perturber interior to TOI-1422\,b; (3) a three-planet model with a perturber between TOI-1422\,b and TOI-1422\,c; and (4) a three-planet model with an external perturber beyond TOI-1422\,c.
For all the configurations we used as fitting parameters: the planetary-to-star mass ratio $M_\mathrm{p}/M_\star$, the period $P_{\rm p}$, the eccentricity $e_{\rm p}$ and the mean longitude $\lambda_{\rm p}$\footnote{$\lambda = \mathcal{M} + \omega + \Omega$, where $\mathcal{M}$ is the mean anomaly, $\omega$ is the argument of periastron (or pericenter), and $\Omega$ is the longitude of the ascending node.}. To improve sampling efficiency and avoid biases at low eccentricities, we adopted the parametrization $[\sqrt{e} \cos \omega, \sqrt{e} \sin \omega]$ \citep{Eastman_2013} instead of fitting the eccentricity $e$ and the argument of the periastron $\omega$ directly. The stellar radius ($R_\star$) and the stellar mass ($M_\star$) were fixed to the values reported by \citealt{Naponiello2022}.
The radii ($R_{\rm p}$) and orbital inclinations of the two transiting planets were fixed to the values derived in Sect.~\ref{sec:globalfit}, while the inclination of the potential third planet was treated as a free parameter of the fit. The longitude of the ascending node was fixed at $\Omega = 180^\circ$ for planet b, while it was allowed to vary $180\pm 10^{\circ}$, for planets c and d, in line with previous works \citep[e.g.,][]{winn_2010,Borsato2014}. For the RV data set, we included a jitter term $\sigma_\mathrm{jitter}$ (parametrized in $\log_2$) and a systemic velocity offset $\gamma_{RV}$ as free parameters in the fit.
All parameters were defined at the reference epoch 2\,458\,745.0 ($\mathrm{BJD_{TDB}}$) and integrated over a baseline of 1910 days, corresponding to the full time span covered by the available photometric and RV observations.

We initially ran \trades{} with \pyde, using 120 walkers for 100\,000 steps. The resulting best-fit solutions were then used as starting points for \emcee, which we ran for 400\,000 steps using 120 walkers and a thinning factor of 100. Following the approach of \citet{Nascimbeni_2024}, we adopted a mixed proposal distribution in \emcee, applying the differential evolution proposal for 80\% of the walkers \citep{nelson_2014} and the snooker differential evolution proposal for the remaining 20\% \citep{terbraak_2008}. 
For each of the four analysis, we chose the steps to discard as burn-in through multiple diagnostics, including the Geweke statistic \citep{geweke_1991}, the Gelman-Rubin test \citep{gelmanrubin_1992}, the autocorrelation function \citep{goodman_weare_2010}, and visual inspection of the chains.

Parameter uncertainties were derived from the marginalized posterior distributions as the 68.27\% Highest Density Intervals (HDIs), corresponding to the regions with the highest posterior probability. We defined the best-fit values as the Maximum A Posteriori (MAP) estimates, constrained to lie within the HDIs, representing the most probable set of parameters given the data and priors.

The best-fit parameters derived from each model, along with their corresponding priors, are reported in Table~\ref{Table:TRADES_results}. The observed-minus-calculated (O–C) diagrams for all four tested configurations are shown in Figs.~\ref{fig:oc_1}–\ref{fig:oc_4}.
Among the scenarios tested, only the third configuration, which includes a third planet located between the orbits of TOI-1422\,b and c, yields planetary masses and eccentricities for both b and c that are consistent with those derived from the global fit (see Table~\ref{tab:2p}).
In this configuration, the additional planet d follows an eccentric orbit located near a 3:2 MMR with TOI-1422\,c and a 5:3 MMR with the inner planet TOI-1422\,b, suggesting that the system may be close to a 2:3:5 resonant chain. Its inferred mass is low ($M_{\rm_d} = 2.89_{-0.82}^{+0.95} \, M_{\oplus}$; $K_{\rm d} = 0.57\pm0.20$\,m\,s$^{-1}$), suggesting that it could have remained undetected in the RV data set (see Sect.~\ref{sec:detection_limits}). 
The other configurations require high eccentricities and very low masses that are incompatible with the global fit constraints. 

We repeated the analyses described in Sect.\,\ref{sec:periodograms} to search for evidence of an additional transiting planet between the orbits of TOI-1422\,b and TOI-1422\,c, but found no significant indication. Additional transit observations of planets b and c are therefore critical to confirm and expand the TTV baseline, enabling a more robust detection and characterization of the potential perturbing companion.

\begin{table*}
\centering
\small
\renewcommand{\arraystretch}{1.3}
\caption{Derived parameters of the TOI-1422 system from the four dynamical scenarios tested.}
\resizebox{\textwidth}{!}{%
\begin{tabular}{l c c c c c c c c}
\hline\hline
\textbf{Configurations} & \multicolumn{2}{c}{2 Planet} & \multicolumn{2}{c}{3 Planet (inner)} & \multicolumn{2}{c}{3 Planet (middle)} & \multicolumn{2}{c}{3 Planet (outer)} \\
\hline
\textbf{Parameter} & Prior & MAP (HDI$\pm1\sigma$) & Prior & MAP (HDI$\pm1\sigma$) & Prior & MAP (HDI$\pm1\sigma$) & Prior & MAP (HDI$\pm1\sigma$) \\
\hline
\emph{TOI-1422\,b} \rule{0pt}{12pt} & & & & & & & & \\
Orbital Period ($P$) [days] & \unif{12.5}{13.5} &  $12.9937_{-0.0014}^{+0.0022}$ &  \unif{12.5}{13.5} & $12.99311_{-0.00310}^{+0.00010}$ & \unif{12.5}{13.5} & $12.9990_{-0.0026}^{+0.0011}$ & \unif{12.5}{13.5} & $12.99295_{-0.00054}^{+0.00278}$ \\
Mass [$M_{\oplus}$] & \unif{0.1}{318} & $0.928_{-0.099}^{+0.807}$ & \unif{0.1}{318} & $1.80_{-0.22}^{+0.62}$ & \unif{0.1}{318} & $8.58_{-0.74}^{+2.60}$ & \unif{0.1}{318} &  $1.32_{-0.44}^{+0.54}$ \\
Eccentricity [deg] & \unif{0}{0.5} &  $0.161_{-0.043}^{+0.078}$ & \unif{0}{0.5} &  $0.048_{-0.030}^{+0.019}$ & \unif{0}{0.5} & $0.040_{-0.035}^{+0.027}$ & \unif{0}{0.5} & $0.142_{-0.016}^{+0.095}$ \\
\hline
\emph{TOI-1422\,c} \rule{0pt}{12pt} & & & & & & & & \\
Orbital Period ($P$) [days] & \unif{33}{35.5} & $34.56493_{-0.00065}^{+0.00139}$ & \unif{33}{35.5} & $34.56575_{-0.00027}^{+0.00296}$ & \unif{33}{35.5} & $34.56791_{-0.00019}^{+0.00265}$  & \unif{33}{35.5} &  $34.56586_{-0.00168}^{+0.00045}$ \\
Mass [$M_{\oplus}$] & \unif{0.1}{318} & $16_{-2}^{+4}$ & \unif{0.1}{318} & $19_{-4}^{+2}$ & \unif{0.1}{318} & $14_{-3}^{+2}$ & \unif{0.1}{318} & $20_{-4}^{+2}$ \\
Eccentricity [deg] & \unif{0}{0.5} & $0.198_{-0.036}^{+0.031}$ & \unif{0}{0.5}  & $0.245_{-0.013}^{+0.042}$  & \unif{0}{0.5} & $0.016_{-0.016}^{+0.029}$ & \unif{0}{0.5} & $0.191_{-0.033}^{+0.030}$ \\
\hline
\emph{TOI-1422\,d} \rule{0pt}{12pt} & & & & & & & & \\
Orbital Period ($P$) [days] &  &  & \unif{1}{11} & $2.189_{-0.043}^{+4.223}$ & \unif{14}{31} & $21.704_{-0.012}^{+0.016}$ & \unif{40}{400} & $365_{-25}^{+35}$ \\
Mass [$M_{\oplus}$] &  &  & \unif{0.1}{318} & $0.63_{-0.54}^{+1.34}$ & \unif{0.1}{318} & $2.89_{-0.82}^{+0.95}$ & \unif{0.1}{318} &  $40_{-25}^{+8}$\\
Eccentricity [deg] &  &  & \unif{0}{0.5} &  $0.12_{-0.12}^{+0.19}$ & \unif{0}{0.5} & $0.1589_{-0.0082}^{+0.0818}$ & \unif{0}{0.5} & $0.414_{-0.155}^{+0.086}$ \\
Inclination [deg] &  &  & \unif{80}{100}& $89_{-6}^{+7}$ & \unif{80}{100}& $91_{-6}^{+6}$ &  \unif{80}{100} & $94_{-12}^{+1}$ \\
\hline
\end{tabular}
}
\label{Table:TRADES_results}
\end{table*}

\subsection{Detection limits}
\label{sec:detection_limits}
We applied the \textit{Algorithm for the Refinement of DEtection limits via N-body stability Threshold} (\texttt{ARDENT}; \citealt{Stalport2025})\footnote{\url{https://github.com/manustalport/ardent}} to compute the dynamical HARPS-N RV detection limits for the TOI-1422 system. Namely, \texttt{ARDENT} incorporates orbital stability constraints in addition to the data-driven limits, ensuring that only dynamically viable unseen candidates are retained.

\begin{figure}
\centering
\includegraphics[width=0.48\textwidth]{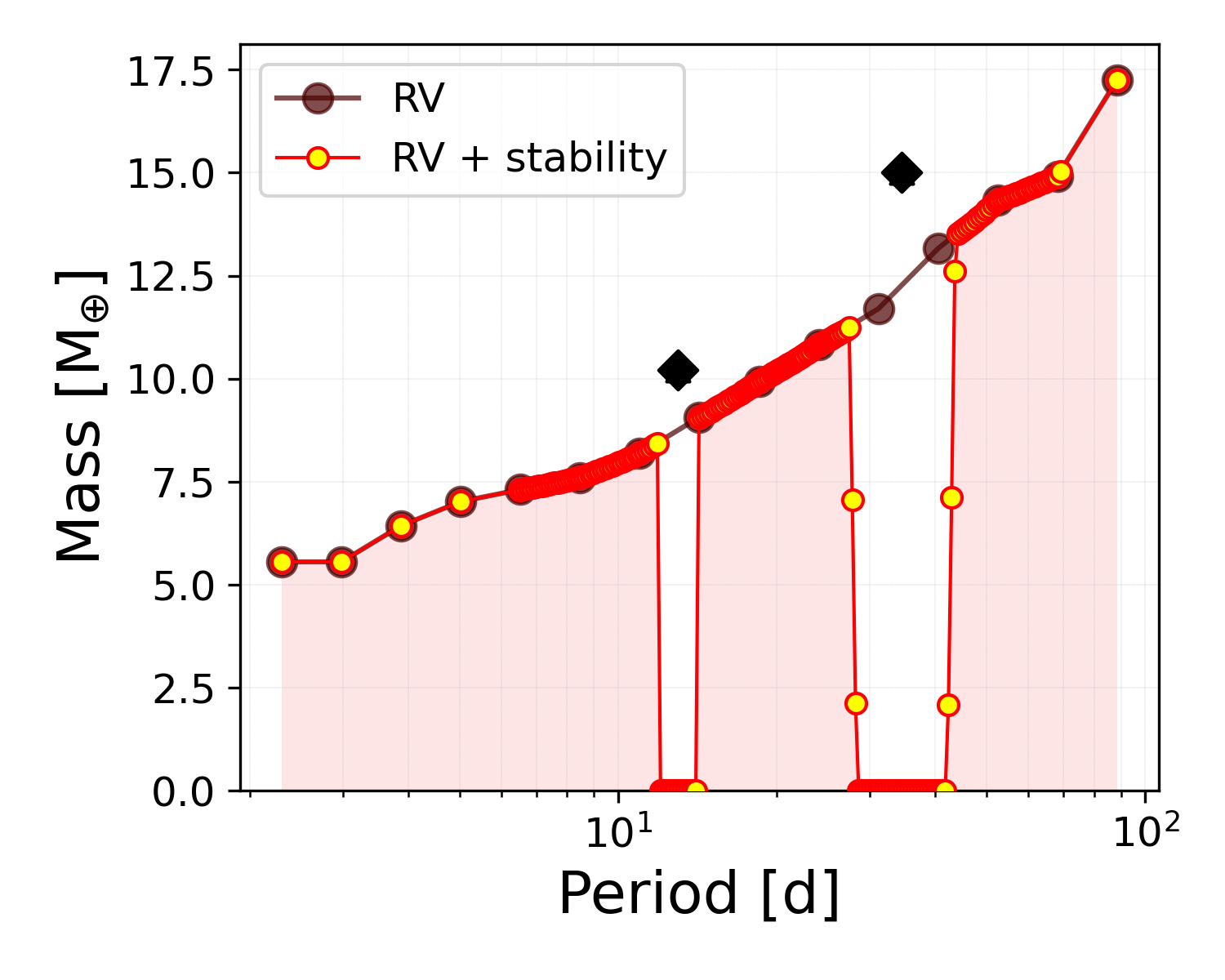}
\caption{RV detection limits in the TOI-1422 system computed with ARDENT. The data-driven detection limits are presented in brown. The dynamical detection limits are illustrated in red-yellow. The known planets TOI-1422 b and c are indicated with the black diamond symbols. The red shaded area depicts the region of the period-mass space where additional candidates could be expected.}\label{fig:DetectionLimits}
\end{figure}

In particular, we performed planet injection–recovery tests in the $P$–$K$ parameter space by sampling 5000 points across the grid. The period of the synthetic planets was drawn from a log-uniform distribution between 2 and 100 days, the RV semi-amplitude $K$ was drawn uniformly around the root-mean-square of the RV uncertainties, the eccentricity was sampled from the beta distribution (following \citealt{Kipping2013}) and the argument of periastron $\omega$ was sampled uniformly between -$\pi$ and $\pi$, while the inclination was fixed to 90 degrees. For each $P$-$K$ point, we carried out ten injection-recovery tests with different injected planets' orbital phases evenly spaced in [-$\pi$ ; $\pi$]. We applied a technique similar to \citet{Bonomo2023} for each test, where RVs at each observed time stamp are drawn from a normal distribution with the mean given by the predicted injected planet signal and the standard deviation given by the measurement uncertainty. The synthetic planet is considered recovered if the most significant signal in the periodogram is within 5$\%$ of the injected period and its FAP is below 1$\%$. The detection rate of each planet in the $P$-$K$ space is given by the number of detections out of the ten tests. These experiments enable the computation of detection limits based on the RV data, namely the data-driven detection limits. In Fig. \ref{fig:DetectionLimits}, we present the 95$\%$ detection limits after converting $K$ into minimal mass (brown curve). 

As a second step, we investigated the dynamical viability of injected planets considering their gravitational interaction with the known ones. The motivation is that real planets must be dynamically viable in a relatively old system such as TOI-1422, the system being unstable otherwise. This additional information, which is based on orbital stability, allows us to push further the detection limits by removing unstable regions of the period-mass space (cf. \citealt{Stalport2025}, for further details). \texttt{ARDENT} computes the updated, dynamical detection limits from the data-driven detection limits and the orbital parameters of the two known planets (fixing here the non-significant orbital eccentricities to zero). We applied a dense period sampling around the known planets by interpolating the data-driven limit curve in 100 period bins inside the intervals [$P_i/2$; $2 P_i$] (where $i$ denoted the known planet index). The orbital stability was estimated for each point on this limit curve, and for lower masses if the former was unstable. Fig. \ref{fig:DetectionLimits} shows the outcome, with the dynamical detection limits in red. 

We found significant room left for an additional third planet candidate in the system. In particular, the instability zones where no planet can exist roughly cover the ranges [$11.5-14.5$] days and [$28.6-42.3$] days. Outside these period windows around the known planets, an additional candidate could exist and remain undetected with the current RV data at hand, especially if $M_p\lesssim5\,M_{\oplus}$ (such as the candidate suggested in Sect.\,\ref{sec:dynamical}).

\section{Compositions}\label{sec:formation}
We analyse the potential composition of TOI-1422\,c using a static inference model based on \citet{dorn_generalized_2017} with updates described in \citet{luo_majority_2024}. 
The model consists of three layers: an iron core, a silicate mantle, and a H$_2$-He-H$_2$O atmosphere. 
We assume an adiabatic temperature profile for the core and mantle and allow for both liquid and solid phases in the two layers.
For liquid iron and iron alloys we use the equation of state (EOS) by \citet{luo_majority_2024}. For solid iron, we use the EOS for hexagonal close packed iron \citep{hakim_new_2018,miozzi_new_2020}. 
For pressures below $\approx 125\,$GPa, the solid mantle mineralogy is modelled using the thermodynamical model \textsc{Perple\_X} \citep{connolly_geodynamic_2009} considering the system of MgO, SiO$_{2}$, and FeO. 
At higher pressures we define the stable minerals \textit{a priori} and use their respective EOS from various sources \citep{hemley_constraints_1992,fischer_equation_2011,faik_equation_2018,luo2023equation}.
The liquid mantle is modelled as a mixture of Mg$_2$SiO$_4$, SiO$_2$ and FeO \citep{melosh_hydrocode_2007,faik_equation_2018,ichikawa_ab_2020,stewart_shock_2020}, and mixed using the additive volume law. 

The H$_2$-He-H$_2$O atmosphere layer is modelled using the analytic description of \citet{guillot_radiative_2010} and consists of an irradiated layer on top of a non-irradiated layer in radiative-convective equilibrium. 
The water mass fraction is given by the metallicity $Z$ and the hydrogen-helium ratio is set to solar. 
The two components of the atmosphere, H$_2$/He and H$_2$O, are again mixed following the additive volume law and using the EOS by \citet{1995_Saumon_EOS} for H$_2$/He and the ANOES EOS \citep{1990_thompson_aneos} for H$_2$O. For each model realization, the planet intrinsic luminosity is calculated following \citep{mordasini_planetary_2020} and is a function of planet mass, atmospheric mass fraction and age, for which we use the estimate of $5.1^{+3.9}_{-3.1}$ Gyrs from \citealt{Naponiello2022}.

For the inference, we use a surrogate model using Polynomial Chaos Kriging \citep{marelli2014uqlab,schobi2015polynomial}. 
In this approach, the global behaviour of the full physical forward model is replaced by the surrogate model leading to a strong decrease in computational cost (De Wringer, in review). For our inference, the surrogate model provides high quality fits with R-squared values (coefficient of determination) of 0.9999 and 0.9998 for planetary mass and radius, respectively. Also, rms errors are well below observational uncertainties 0.003 and 0.01 for planetary mass and radius, respectively. Those errors of the model uncertainty are accounted for in the likelihood function.
The prior parameter distribution is listed in Table \ref{tab:comppriors} and the results of the inference model are summarized in Table \ref{tab:pposterior}. 
In order to explain the observed radius of TOI-1422\,c a significant atmospheric mass fraction is needed. With the assumption of an Earth-like core-mantle ratio of 0.325:0.675, we find an atmospheric mass of $M_\mathrm{atm}=0.51^{+0.46}_{-0.26}\ M_{\oplus}$, and a poorly constrained metallicity of $Z=0.47 ^{+0.31}_{-0.29}$.
We repeat the composition analysis for TOI-1422\,b and find that planet b likely possesses a larger atmosphere with a lower metallicity than planet c, which directly stems from its lower density.

\begin{table}
\centering
\caption{Inference results for the internal compositions of TOI-1422\,c. Stated errors represent 84th and 16th-percentiles.}\label{tab:pposterior}
\renewcommand{\arraystretch}{1.2}
\begin{tabular}{lcc}
    \hline\hline
     Parameters & TOI-1422\,c & TOI-1422\,b \\
    \hline \\[-6pt]%
    log10($M_\mathrm{atm} \: (M_\oplus)$) & $-0.29^{+0.28}_{-0.3} $ & $0.07^{+0.24}_{-0.25}$\\
    $M_\mathrm{core+mantle} \: (M_\oplus)$ & $13.39^{+2.53}_{-2.53}$ & $8.32^{+1.94}_{-1.68}$\\
    $Z_\mathrm{env}$ & $0.47 ^{+0.31}_{-0.29} $ & $0.21^{+0.19}_{-0.14}$\\
    \bottomrule
\end{tabular}
\end{table}

\section{Discussion}\label{sec:discussion}

The most remarkable feature of the TOI-1422 system is its anti-ordered configuration \citep{Mishra2023}, defined by the relative properties of its two known planets. Although the current mass uncertainties make the two planets compatible within $1.25\sigma$, their nominal values suggest an inversion of the commonly observed mass--radius ordering. In contrast to the widely observed trend in multiplanet systems, where planet mass and radius tend to decrease with orbital period (or at least density remains constant or decreases; \citealt{Weiss2014}), this system exhibits the opposite pattern: the outer planet, TOI-1422\,c ($P_{\rm c} > P_{\rm b}$), is more massive ($M_{\rm c} > M_{\rm b}$) yet smaller than the inner planet ($R_{\rm c} < R_{\rm b}$), TOI-1422\,b. Such anti-ordered or ``mass/density-inverted'' architectures are rarely observed, despite some population synthesis models predicting a frequency of $\lesssim$8\% \citep{Mishra2023}. This discrepancy suggests that additional processes beyond simple in-situ core accretion must have influenced the formation and evolution of this system.

\begin{figure}
\centering
\includegraphics[width=0.48\textwidth]{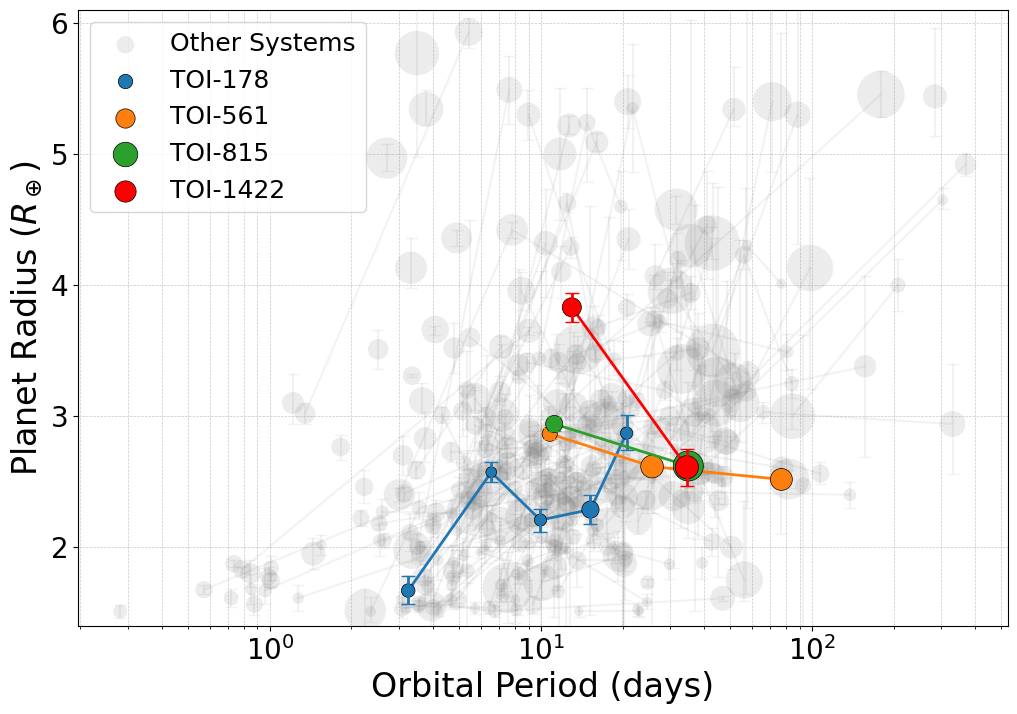}
\caption{Multiplanet systems with at least two planets of radii $1.5\,R_\oplus<R_p<6\,R_\oplus$, as taken from the NASA Exoplanet Database. The systems for which there's at least one outer planet that is both more massive and smaller than the inner planet (with significance $\geq1\sigma$) are coloured and labelled.}\label{fig:antiorder}
\end{figure}

To place our findings in a broader context, we searched the NASA Exoplanet Archive for confirmed multiplanet systems where at least one outer planet is both more massive and smaller than an inner planet (with differences significant at $>1\sigma$) within the radius range 1.5--6~$R_\oplus$, corresponding to super-Earths, sub-Neptunes, and Neptune-type planets. We found only three comparable systems: TOI-178 \citep{Leleu2021}, TOI-561 \citep{Lacedelli2021, Lacedelli2022, Piotto2024}, and TOI-815 \citep{Psaridi2024}. However, in these cases the anti-ordering involves only sub-Neptunes (refer to Fig.\,\ref{fig:antiorder}), whereas in TOI-1422 the inner planet is Neptune-sized.

The observed density contrast between TOI-1422\,b and TOI-1422\,c may point to divergent evolutionary pathways. One possibility involves late-stage giant impacts: a catastrophic collision on planet~c could have stripped its primordial envelope, leaving behind a denser, core-dominated remnant \citep{Inamdar2015,Liu2015}. Alternatively, formation location and migration history may have played key roles \citep{Izidoro2015,Bitsch2019}. Planet~c could have formed in a disk region with an enhanced solid-to-gas ratio or followed a migration path that restricted its access to the natal gas reservoir, resulting in a higher CMF.

Moreover, it has recently been proposed that sub-Neptunes in resonant chains tend to be less dense \citep{Leleu2024}, although this trend may arise from biases between RV and TTV analyses. Depending on the location of the third planet responsible for the TTVs observed on planet b, this system could serve as a valuable test case for exploring that hypothesis. At present, since TOI-1422\,c is not in a known resonance with any other planet, its high density is consistent with the idea that sub-Neptunes outside resonant configurations tend to be denser than those within resonant chains.

\section{Conclusions}\label{sec:conclusions}

Thanks to two new TESS sectors (57 and 84), an expanded RV dataset, and new RV reduction and analysis, we have confirmed the planet candidate TOI-1422\,c. This planet orbits with a period longer than previously suggested ($P_{\rm orb}=34.563$\,d instead of $\sim29$\,d) and, despite being a sub-Neptune, it is more massive than the Neptune-sized sibling TOI-1422\,b ($M_{\rm c}=14\pm3\,M_{\oplus}$ vs. $M_{\rm b}=9.5^{+2.0}_{-1.9}\,M_{\oplus}$), making TOI-1422 a rare anti-ordered system. The two mass estimates are consistent within $1.25\sigma$, yet the comparison highlights that TOI-1422\,c is significantly denser than TOI-1422\,b ($\rho_{\rm c}=4.3^{+1.3}_{-1.0}$ vs. $\rho_{\rm b}=0.93^{+0.21}_{-0.20}$). In addition, our analysis of TOI-1422\,b transit times revealed significant TTVs. Such variations are not caused by gravitational perturbations from TOI-1422\,c. 
Instead, they suggest the presence of an unseen planetary companion that may even be orbiting between them ($P_{\rm d} \approx 21$ days). The confirmation of this possibly low-mass sibling candidate ($M_{\rm d}\lesssim4\,M_{\oplus}$) will require additional photometric observations and extreme-precision ($\lesssim1$\,ms$^{-1}$) RV measurements.

The large density contrast between the inner Neptune and the outer sub-Neptune could point to different formation or evolutionary pathways. A violent history (e.g. giant impacts or strong dynamical instabilities) might provide a possible explanation, whereas a configuration close to a three planet MMR would instead favour a more quiescent dynamical evolution. Atmospheric characterization could help discriminate between these scenarios: a super-solar enrichment in TOI-1422\,c could indicate formation beyond the ice line followed by inward migration, although alternative explanations, such as envelope enrichment through collisions or accretion of metal-rich material, cannot be excluded.

\section*{Acknowledgements}
The work is based on observations made with the Italian Telescopio Nazionale Galileo (TNG) operated on the island of La Palma by the Fundacion Galileo Galilei of the INAF (Istituto Nazionale di Astrofisica) at the Spanish Observatorio del Roque de los Muchachos of the Instituto de Astrofisica de Canarias within the program ID AOT48TAC\textunderscore48.
This work includes data collected with the TESS mission, obtained from the MAST data archive at the Space Telescope Science Institute (STScI). Funding for the TESS mission is provided by the NASA Explorer Program. STScI is operated by the Association of Universities for Research in Astronomy, Inc., under the NASA contract NAS 5–26555. The authors acknowledge the use of public TESS data from pipelines at the TESS Science Office and at the TESS Science Processing Operations Center. Resources supporting this work were provided by the NASA High-End Computing (HEC) Program through the NASA Advanced Supercomputing (NAS) Division at Ames Research Center for the production of the SPOC data products. 
This research has used the Exoplanet Follow-up Observation Program (ExoFOP; DOI: 10.26134/ExoFOP5) website, which is operated by the California Institute of Technology, under contract with the National Aeronautics and Space Administration under the Exoplanet Exploration Program.
This research has used the NASA Exoplanet Archive, which is operated by the California Institute of Technology, under contract with the National Aeronautics and Space Administration under the Exoplanet Exploration Program.
L.N. acknowledges financial contribution from the INAF Large Grant 2023 ``EXODEMO''. L.M. acknowledges the financial contribution from the PRIN MUR 2022 project 2022J4H55R. PLe acknowledges that this publication was produced while attending the PhD program in Space Science and Technology at the University of Trento, Cycle XXXVIII, with the support of a scholarship co-financed by the Ministerial Decree no. 351 of 9th April 2022, based on the NRRP - funded by the European Union - NextGenerationEU - Mission 4 ``Education and Research'', Component 2 ``From Research to Business'', Investment 3.3 -- CUP E63C22001340001.
M.S. thanks the Belgian Federal Science Policy Office (BELSPO) for the provision of financial support in the framework of the PRODEX Programme of the European Space Agency (ESA) under contract number C4000140754.
CD acknowledges support from the Swiss National Science Foundation under grant TMSGI2\_211313. Parts of this work has been carried out within the framework of the NCCR PlanetS supported by the Swiss National Science Foundation under grants 51NF40\_182901 and 51NF40\_205606.
The authors acknowledge the support of the INAF Guest Observer Grant (Normal) ``ArMS: the Ariel Masses Survey Large Program at the TNG'', according to the INAF Fundamental Astrophysics funding scheme.

\section*{Data Availability}
 
The photometric data used in this manuscript is available on the MAST, while the entire RV dataset has been included in Table\,\ref{tab:RVs}.



\bibliographystyle{mnras}
\bibliography{1422c} 




\clearpage
\appendix

\section{Additional figures}

\begin{figure*}
\centering
\includegraphics[width=0.6\textwidth]{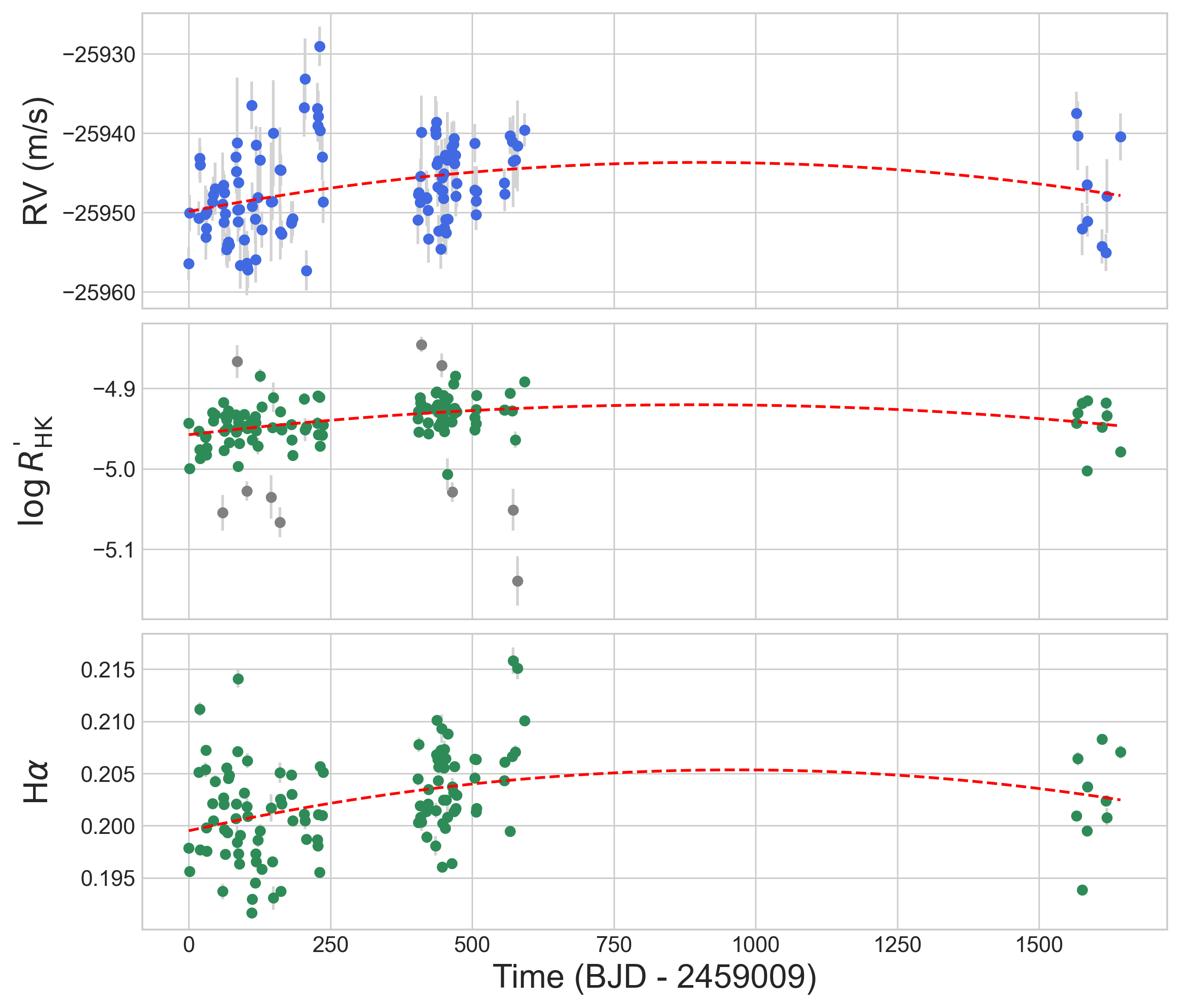}
\caption{RV timeseries along with those of the chromospheric index $\log{R^{\prime}_{\rm HK}}$, the H$\alpha$ line, and the Na\,{\tiny I} line. The dashed red line shows a simple quadratic fit to each dataset, while the gray circles are points that deviate more than 3$\sigma$ from the median and have not been considered for the fit.}\label{fig:long_trend}
\end{figure*}

\begin{figure*}
\centering
\includegraphics[width=0.6\textwidth]{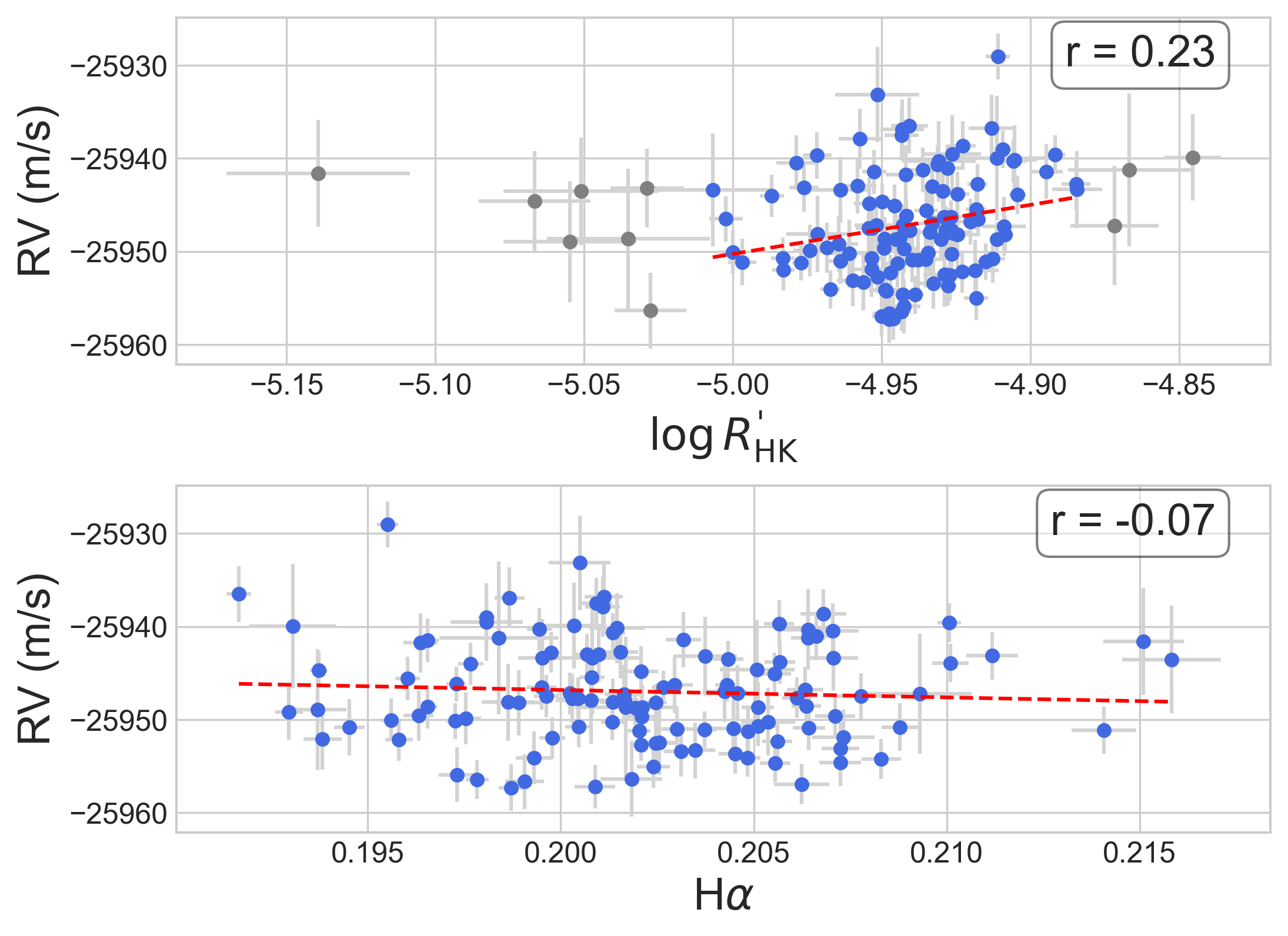}
\caption{Correlation between RVs and stellar activity indicators. The red dashed line represents the best linear fit to the data, and the corresponding Pearson correlation coefficient (r) is reported in the top-right corner of each panel. The gray circles are points that deviate more than 3$\sigma$ from the median and have not been considered.}\label{fig:crosscorr}
\end{figure*}

\begin{figure*}
\centering
\includegraphics[width=0.6\textwidth]{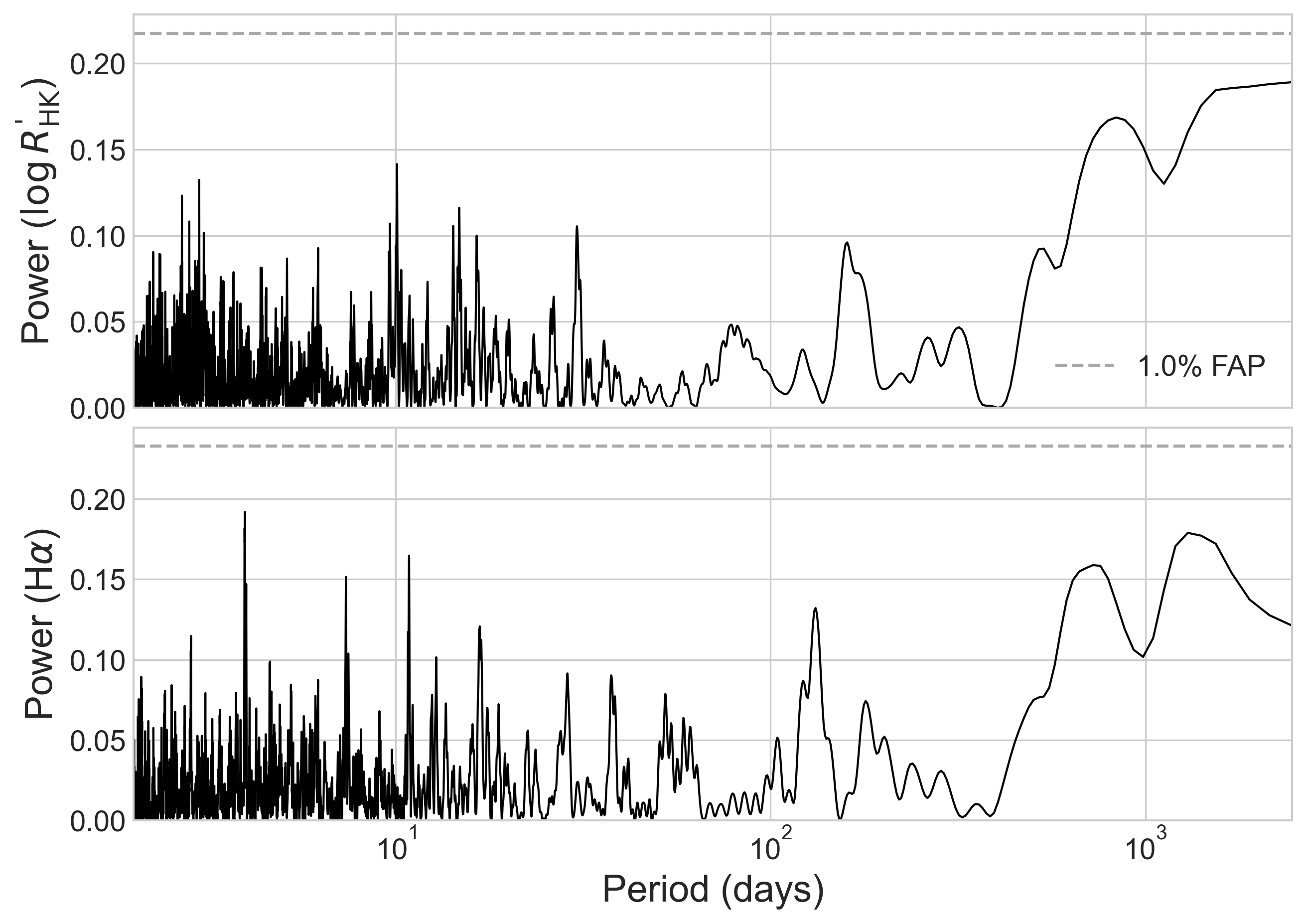}
\caption{GLS periodograms of the $\log{R^{\prime}_{\rm HK}}$ and H$\alpha$ indices. No strong activity peak is revealed, however both show tentative long-period signals.}\label{fig:activity}
\end{figure*}

\begin{figure*}
\centering
\includegraphics[width=1\textwidth]{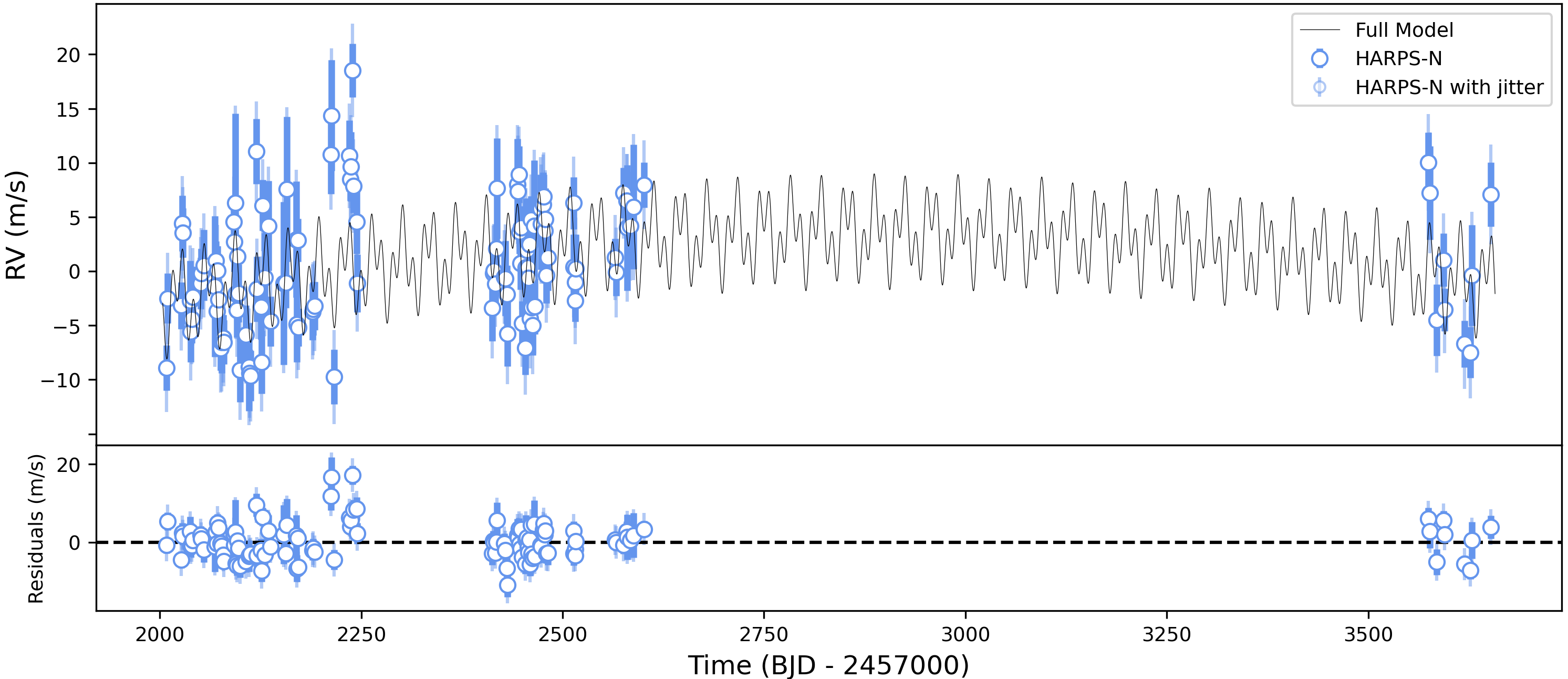}
\caption{The HARPS-N RV dataset of TOI-1422 in blue, along with the global model derived in this paper (top panel) and its residuals (bottom panel).}\label{fig:fullRV}
\end{figure*}

\begin{figure*}
\centering
\includegraphics[width=0.5\textwidth]{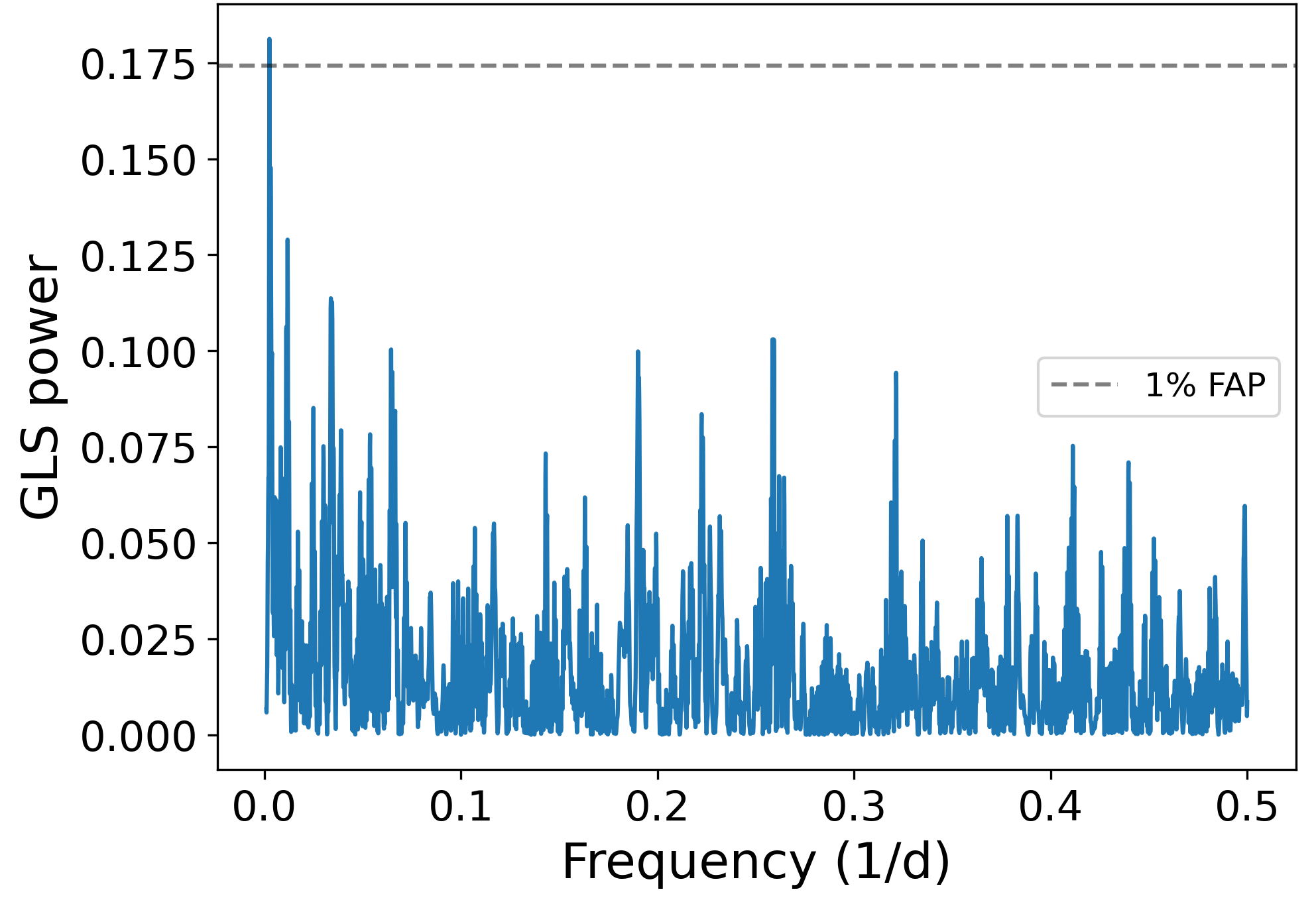}
\caption{GLS periodogram of the RV residuals from the adopted model. The horizontal dashed line marks the 1\% FAP.}\label{fig:periodogram_2p}
\end{figure*}

\begin{figure*}
\centering
\includegraphics[width=0.8\textwidth]{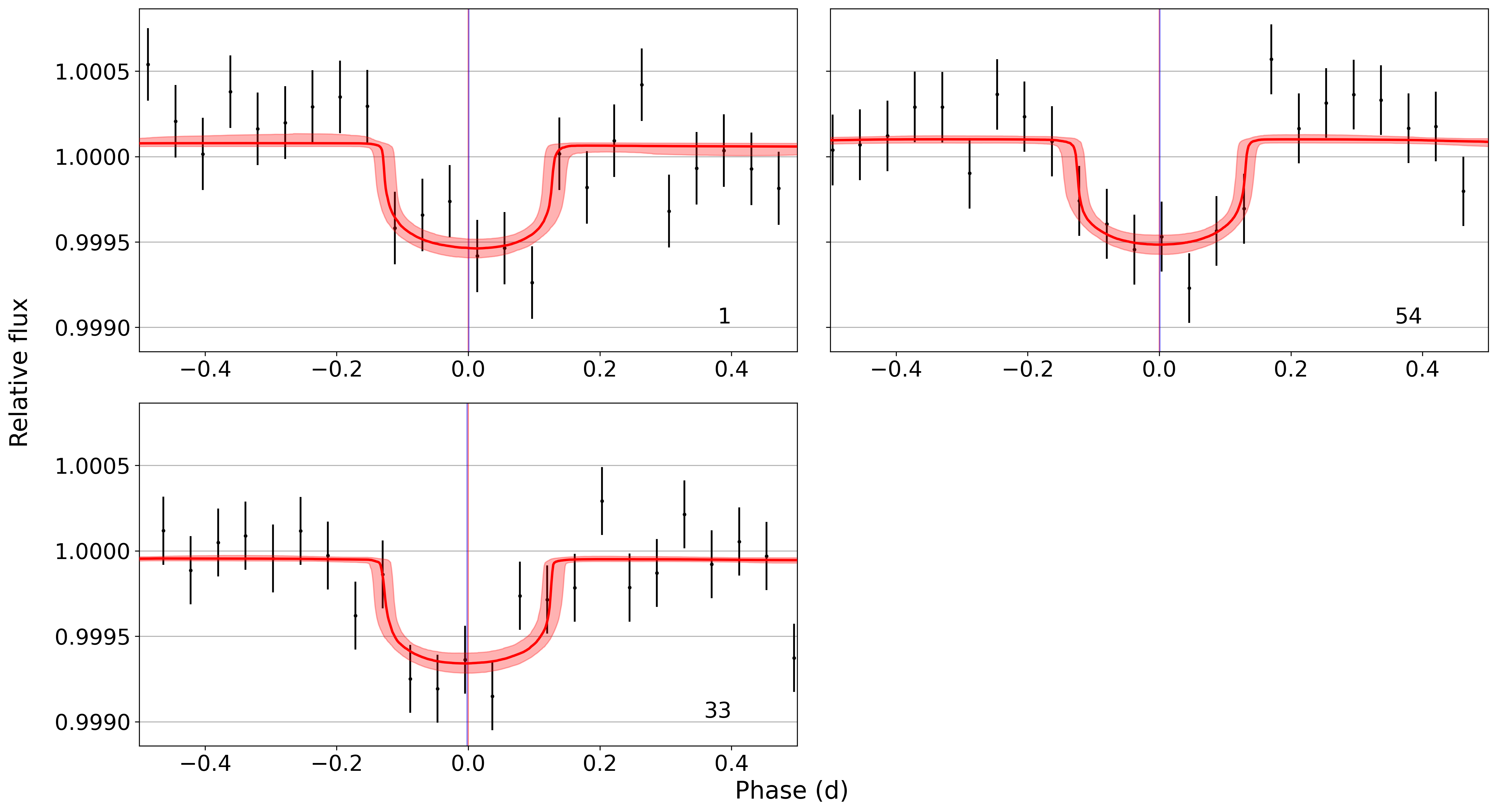}
\caption{The three transits of TOI-1422\,c fitted independently. The center of the transits are indicated by the purple vertical lines.}\label{fig:TTV_planetc}
\end{figure*}

\begin{figure}
    \centering
    \includegraphics[width=1\linewidth]{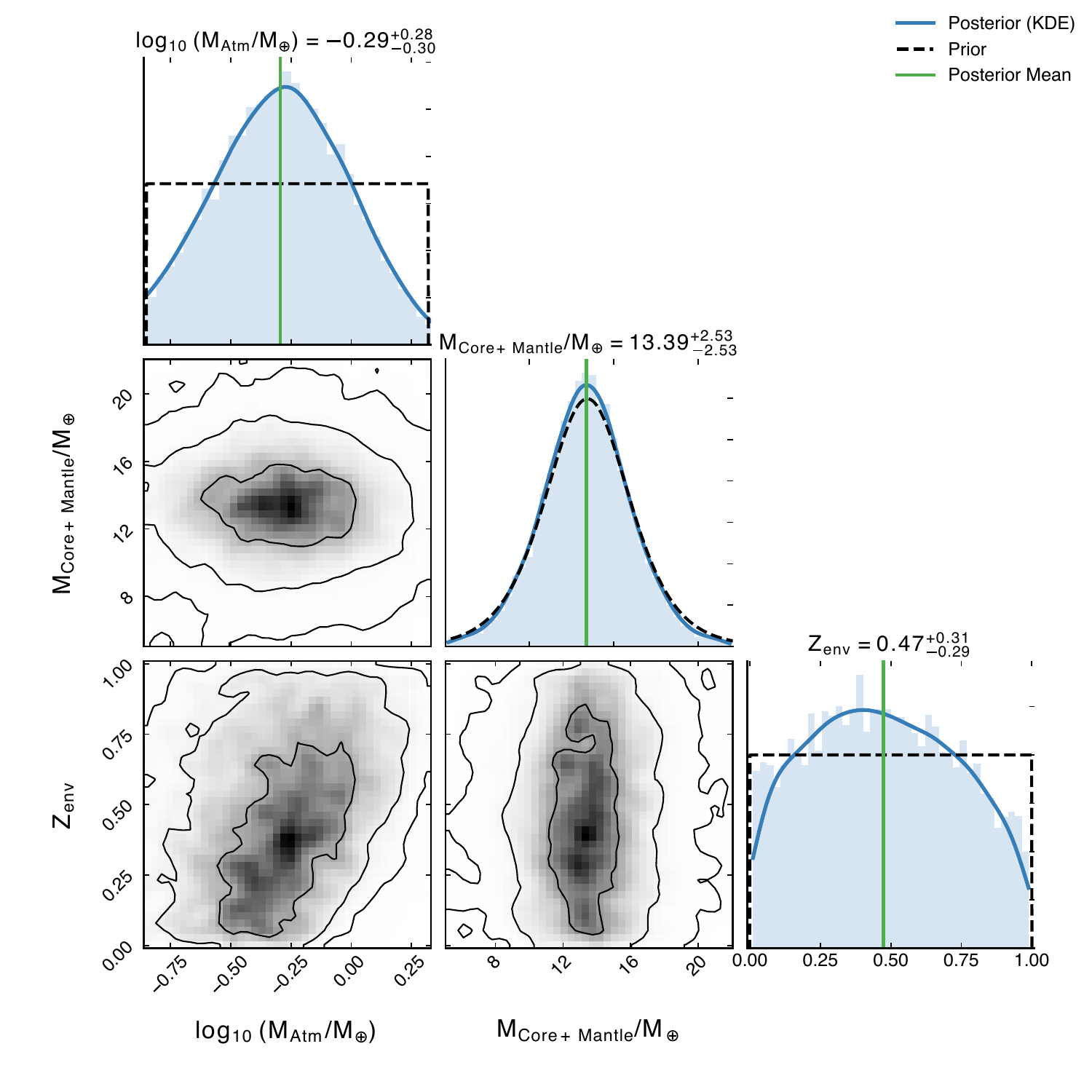}
    \caption{1-D and 2-D Posterior distribution of interior properties of TOI-1422\,c}
    \label{fig:TOI1422ccomp}
\end{figure}

\begin{figure*}
    \centering
    \begin{subfigure}[b]{0.48\textwidth}
        \includegraphics[width=\textwidth,keepaspectratio]{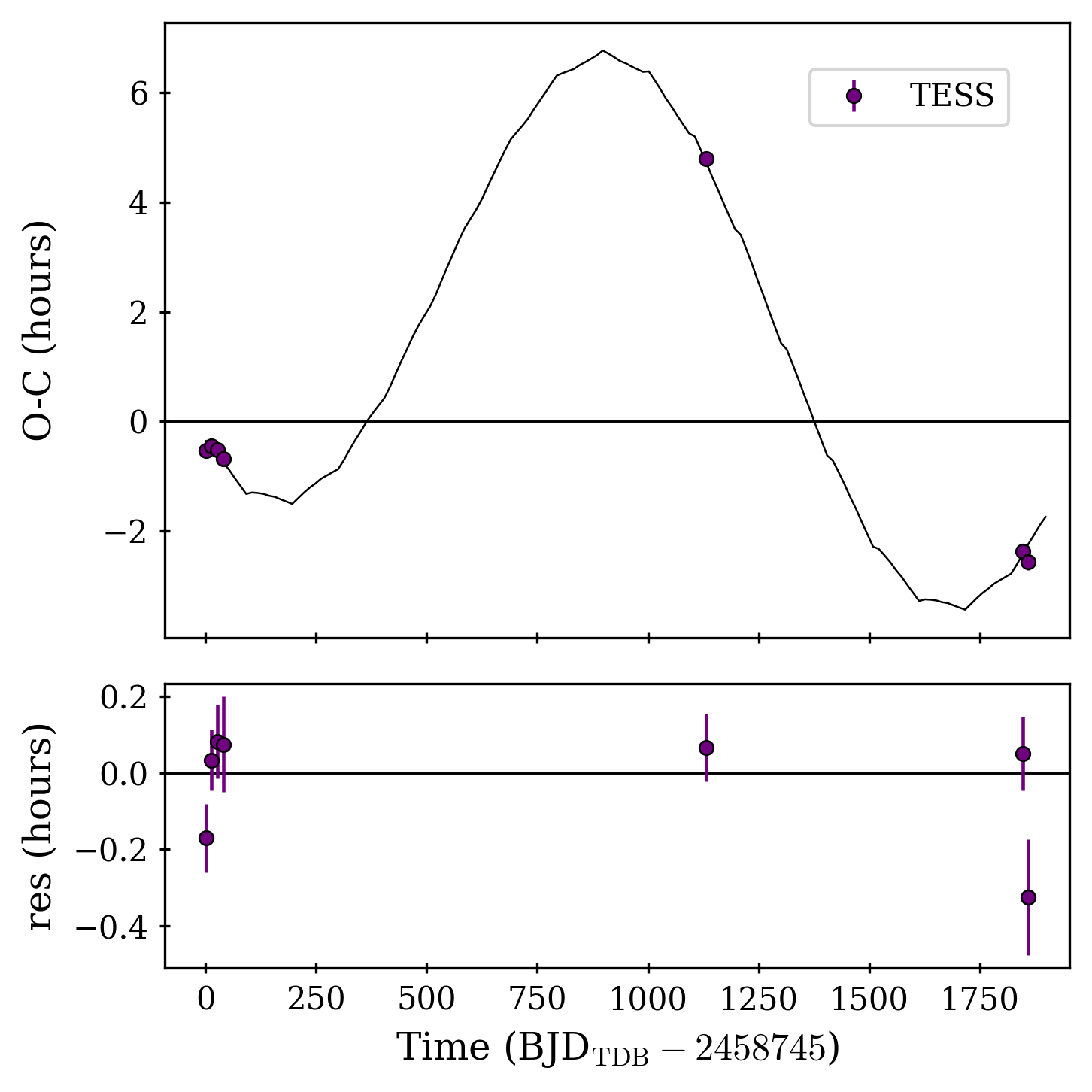}
    \end{subfigure}
    \hfill
    \begin{subfigure}[b]{0.48\textwidth}
        \includegraphics[width=\textwidth,keepaspectratio]{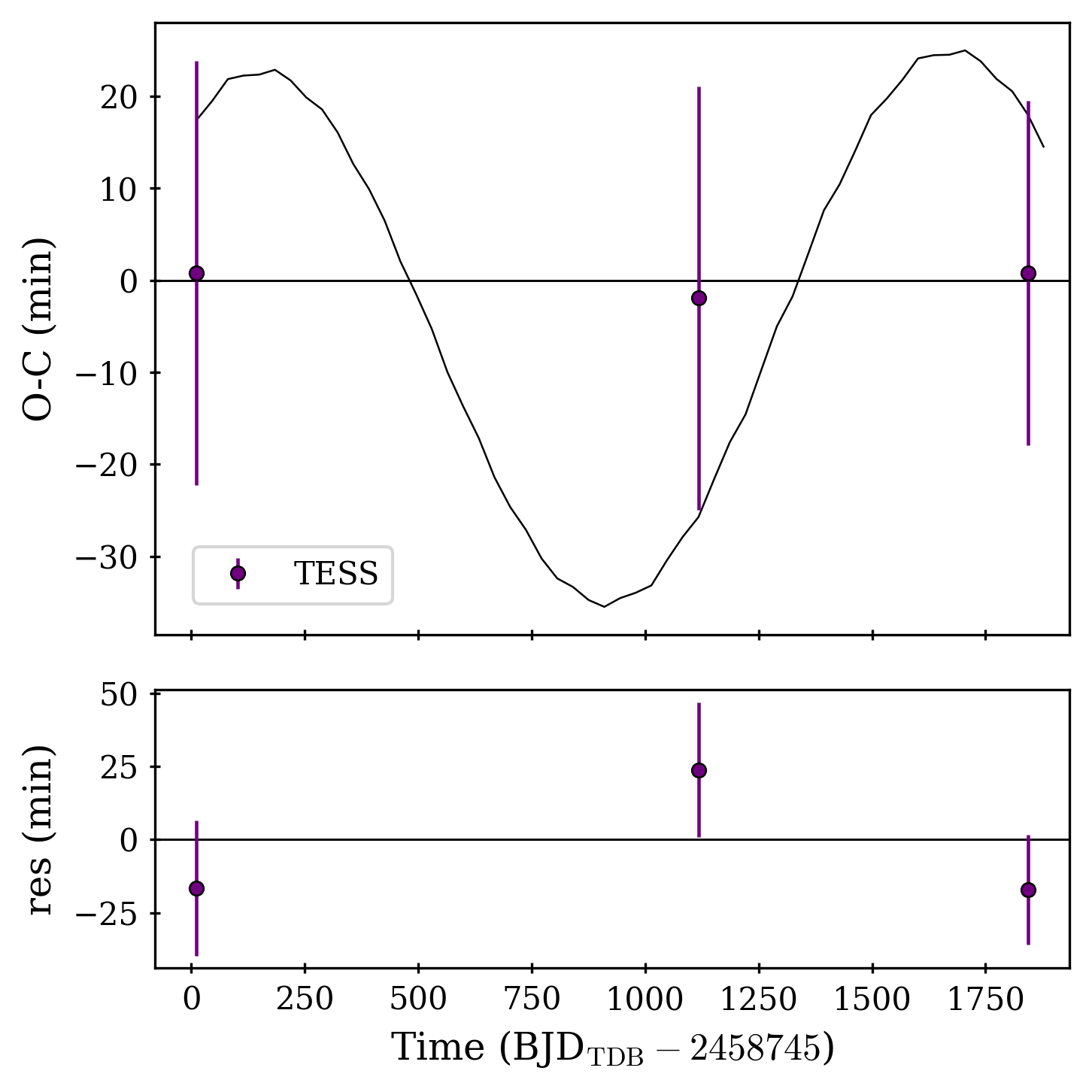}
    \end{subfigure}
    \caption{O-C diagrams of TOI-1422\,b  (left) \&\,c (right) of configuration (1). Upper panel: The best-fit \trades{} model is shown as a black line. Lower panel: Residuals with respect to the best-fit model.}
    \label{fig:oc_1}
\end{figure*}

\begin{figure*}
    \centering
    \begin{subfigure}[b]{0.48\textwidth}
        \includegraphics[width=\textwidth,keepaspectratio]{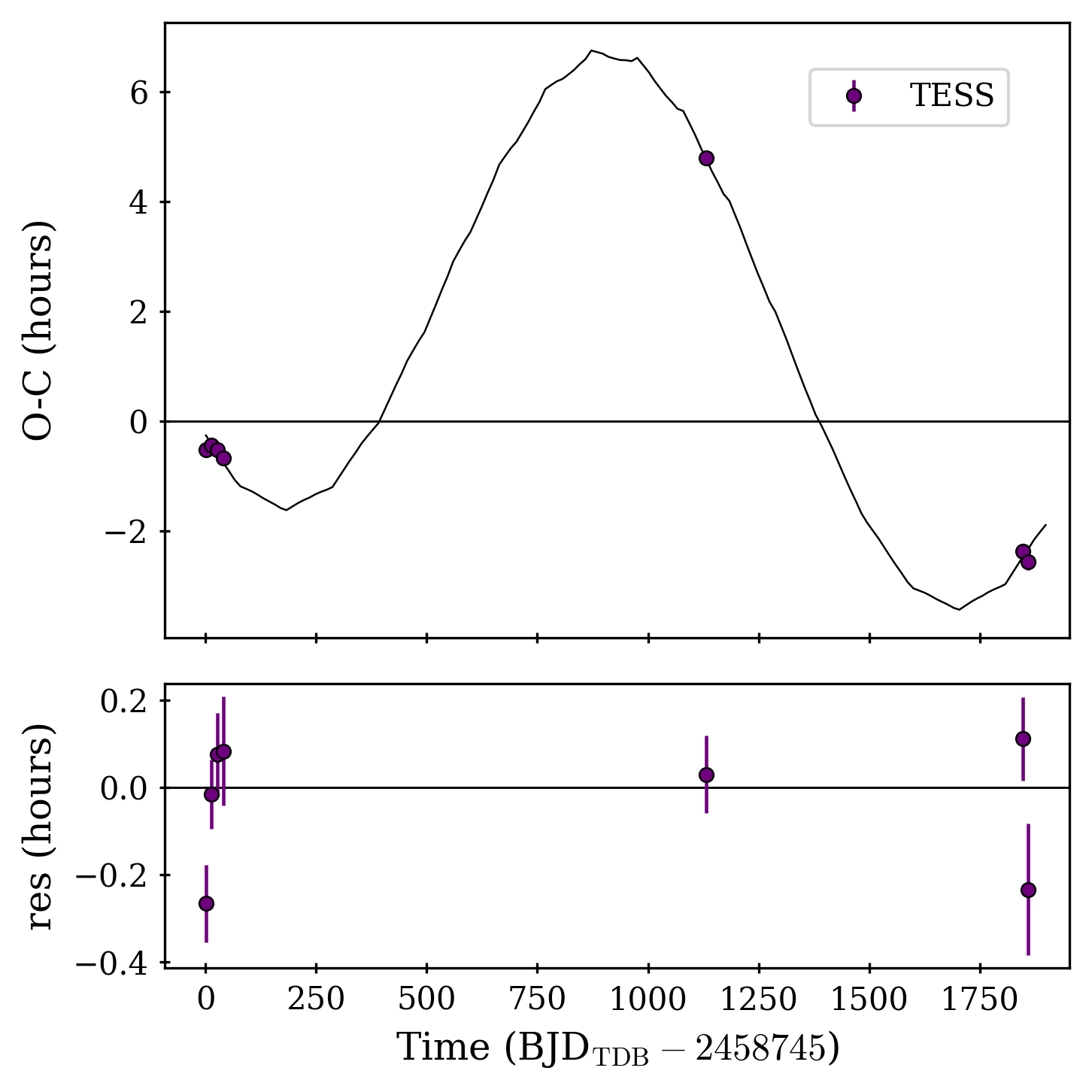}
    \end{subfigure}
    \hfill
    \begin{subfigure}[b]{0.48\textwidth}
        \includegraphics[width=\textwidth,keepaspectratio]{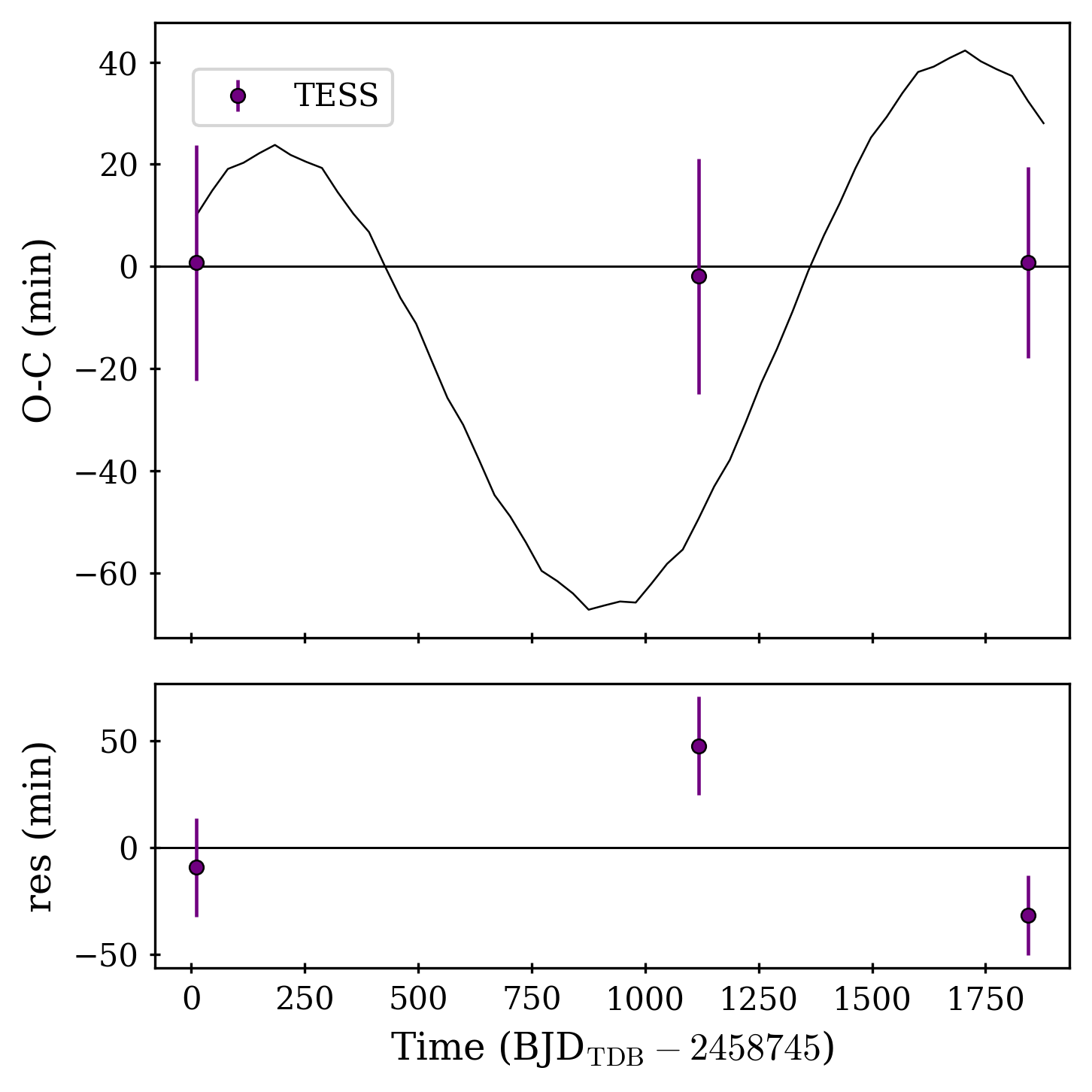}
    \end{subfigure}
    \caption{O-C diagrams of TOI-1422\,b  (left) \&\,c (right) of configuration (2). Upper panel: The best-fit \trades{} model is shown as a black line. Lower panel: Residuals with respect to the best-fit model.}
    \label{fig:oc_2}
\end{figure*}

\begin{figure*}
    \centering
    \begin{subfigure}[b]{0.48\textwidth}
        \includegraphics[width=\textwidth,keepaspectratio]{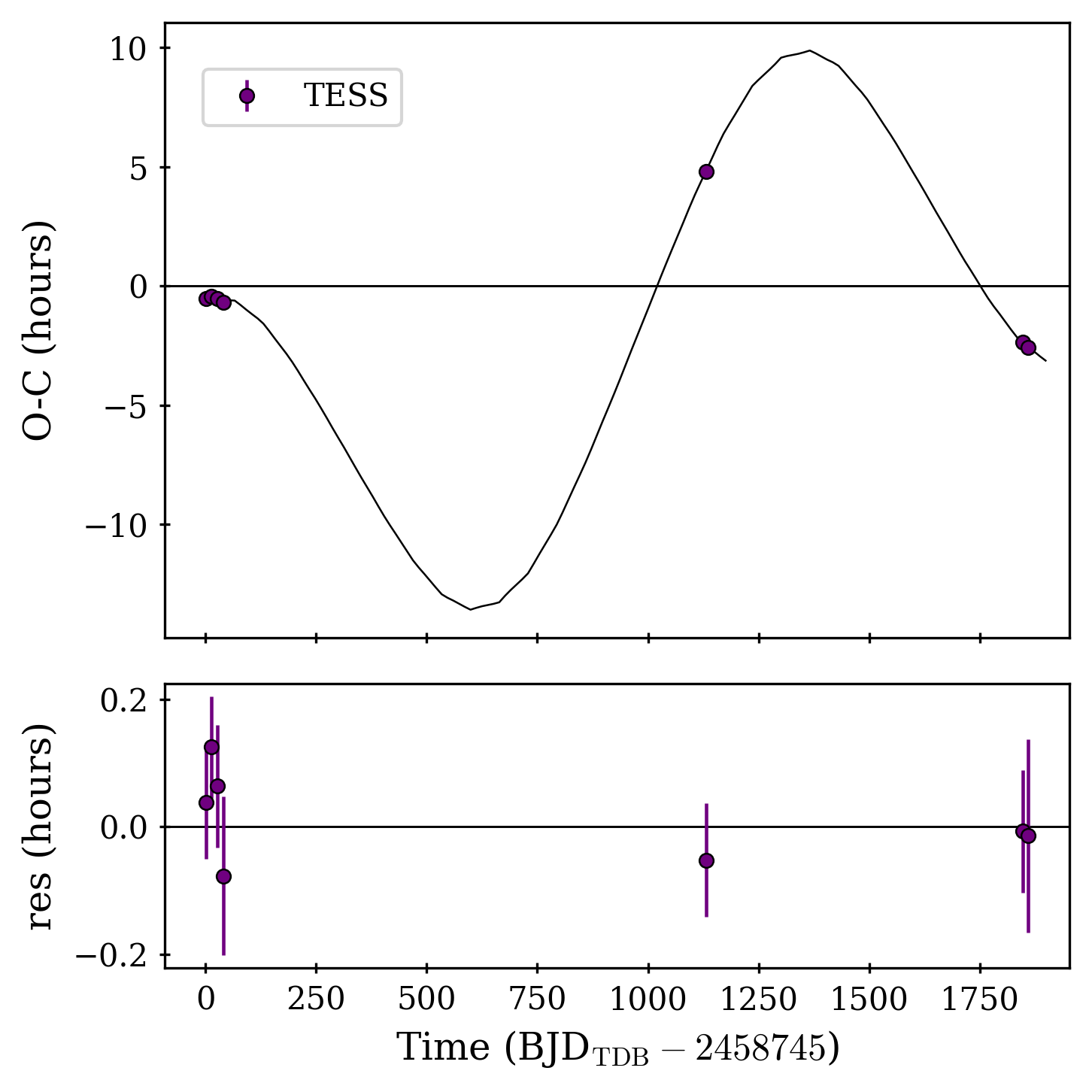}
    \end{subfigure}
    \hfill
    \begin{subfigure}[b]{0.48\textwidth}
        \includegraphics[width=\textwidth,keepaspectratio]{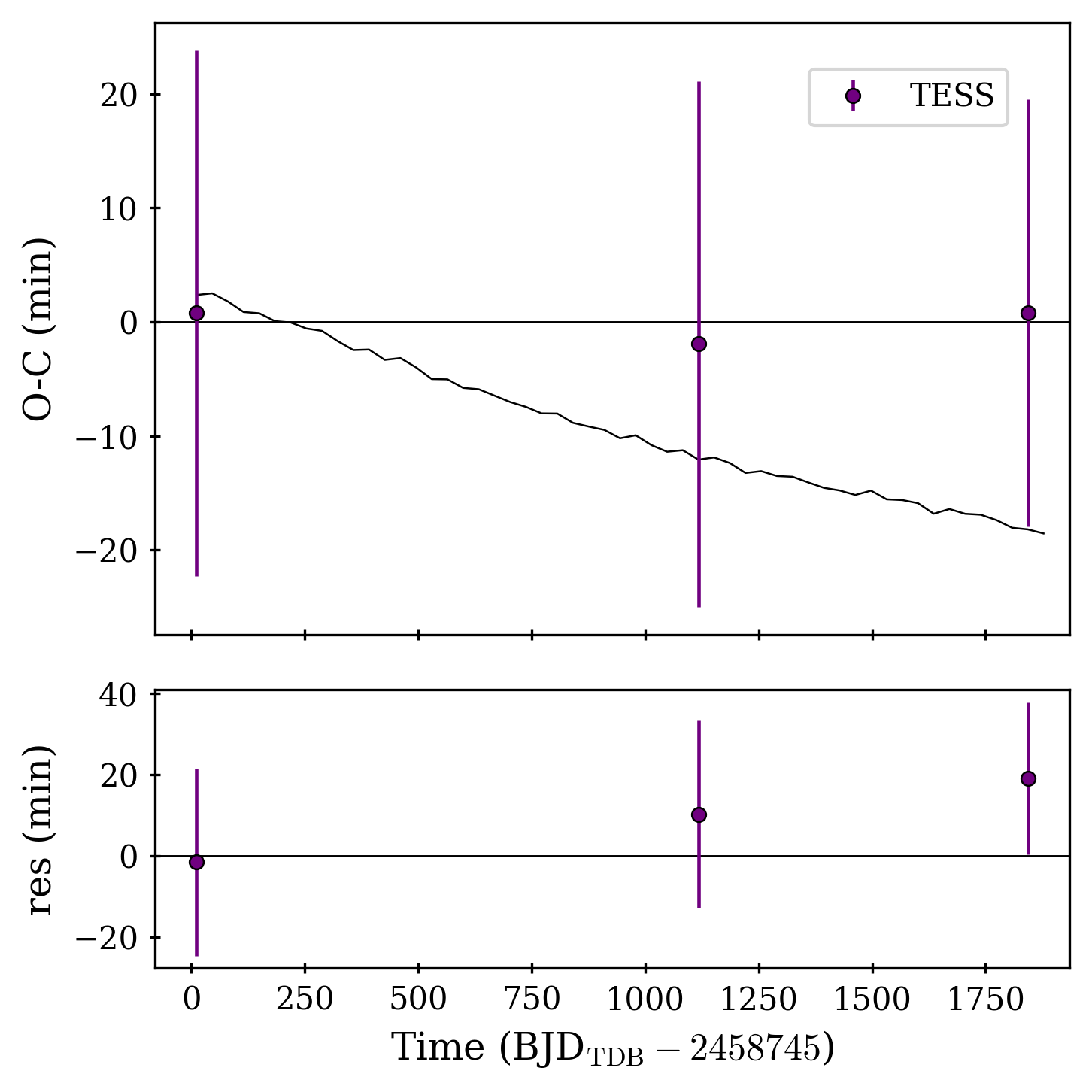}
    \end{subfigure}
    \caption{O-C diagrams of TOI-1422\,b  (left) \&\,c (right) of configuration (3). Upper panel: The best-fit \trades{} model is shown as a black line. Lower panel: Residuals with respect to the best-fit model.}
    \label{fig:oc_3}
\end{figure*}

\begin{figure*}
    \centering
    \begin{subfigure}[b]{0.48\textwidth}
        \includegraphics[width=\textwidth,keepaspectratio]{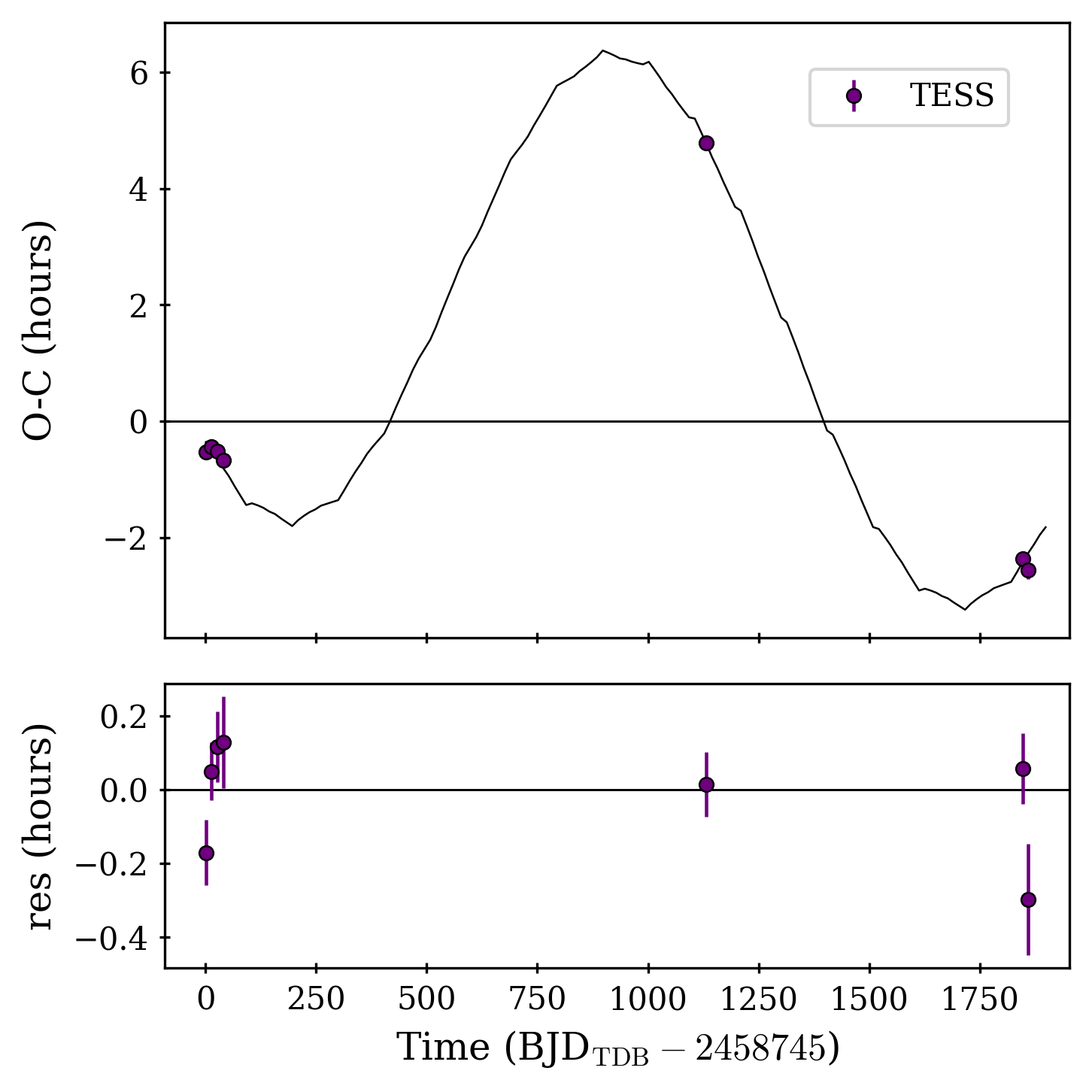}
    \end{subfigure}
    \hfill
    \begin{subfigure}[b]{0.48\textwidth}
        \includegraphics[width=\textwidth,keepaspectratio]{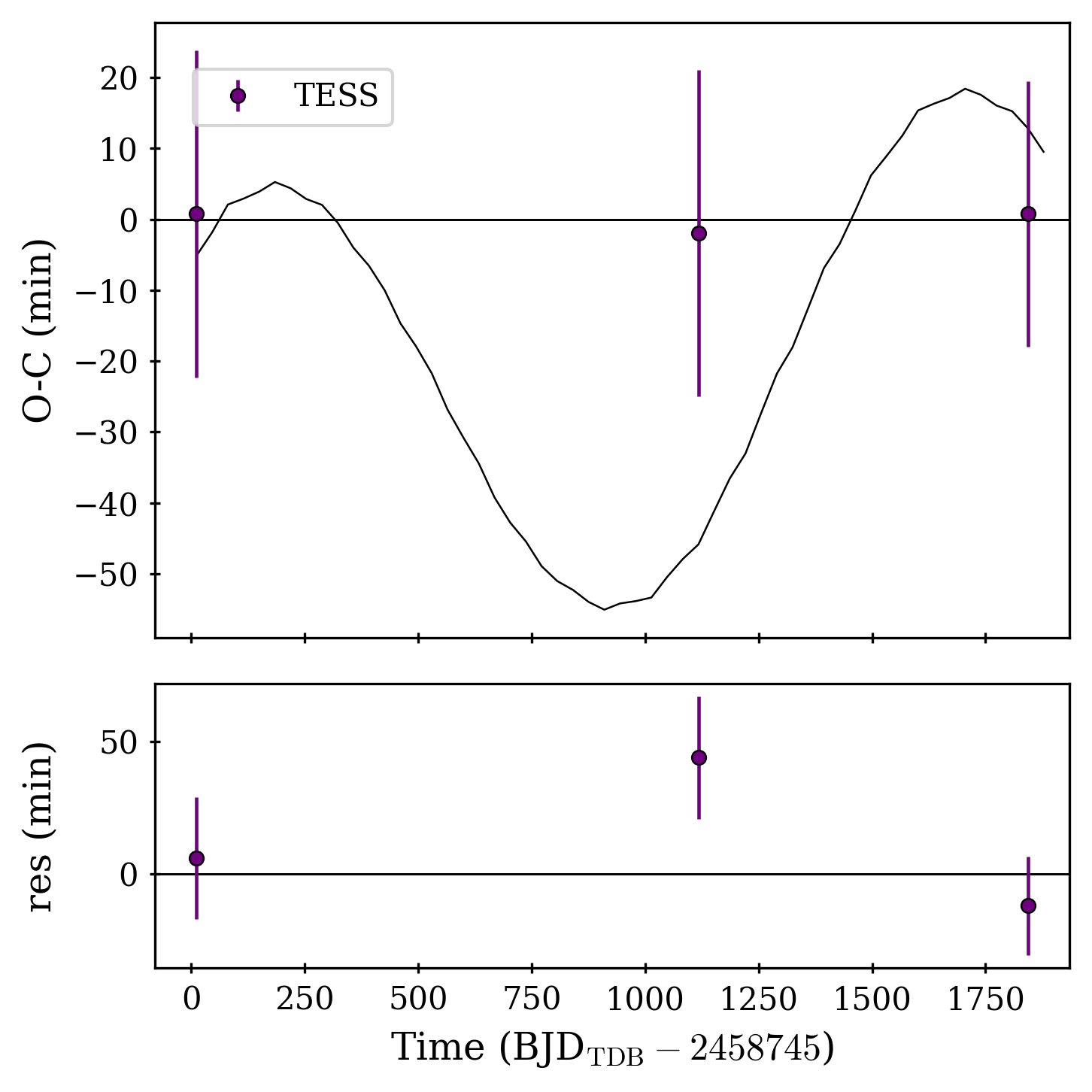}
    \end{subfigure}
    \caption{O-C diagrams of TOI-1422\,b  (left) \&\,c (right) of configuration (4). Upper panel: The best-fit \trades{} model is shown as a black line. Lower panel: Residuals with respect to the best-fit model.}
    \label{fig:oc_4}
\end{figure*}

\section{Additional tables}

\begin{table*}
\centering
\caption[]{HARPS-N RV data points and activity indexes of TOI-1422, including the Full Width at Half Maximum (FWHM), the Contrast and the Bisector inverse span (BIS) of the cross correlation function.}\label{tab:RVs}
{\tiny\renewcommand{\arraystretch}{.8}
\resizebox{!}{.4\paperheight}{%
\begin{tabular}{llllllllllllllll}
    \toprule
    $\mathrm{BJD_{\textsf{UTC}}}$ & RV & $\pm1\upsigma_{\textsf{RV}}$ & FWHM & FWHM err & Contrast & Cont. err & BIS & BIS err & $I_{\rm H\upalpha}$ & $I_{\rm H\upalpha}$ err & $I_{\rm Na~{I}}$ & $I_{\rm Na~{I}}$ err & $\log R^{\prime}_{\rm HK}$ & $\log R^{\prime}_{\rm HK}$ err \\ 
    $-2457000$\,[days] & \multicolumn{2}{c}{$\mathrm{~~[m\,s^{-1}]}$} & --& --& -- & --& --& --& --& --& --& -- & -- & -- \rule[-0.8ex]{0pt}{0pt} \\ 
    \hline \\
2008.69430381 & -25956.44 & 2.08 & 7.2146 & 0.0042 & 51.788 & 0.030 & -0.0488 & 0.0042 & 0.19783 & 0.00030 & 0.33509 & 0.00021 & -4.9432 & 0.0032 \\
2009.71241443 & -25950.05 & 2.31 & 7.2199 & 0.0046 & 51.770 & 0.033 & -0.0390 & 0.0046 & 0.19560 & 0.00026 & 0.32484 & 0.00023 & -4.9999 & 0.0044 \\
2026.66268375 & -25950.71 & 2.14 & 7.2119 & 0.0043 & 51.869 & 0.031 & -0.0432 & 0.0043 & 0.20511 & 0.00045 & 0.34297 & 0.00025 & -4.9532 & 0.0035 \\
2027.72143979 & -25943.15 & 2.57 & 7.2173 & 0.0051 & 51.747 & 0.037 & -0.0461 & 0.0051 & 0.21118 & 0.00067 & 0.34333 & 0.00035 & -4.9760 & 0.0049 \\
2028.69382541 & -25944.00 & 2.27 & 7.2223 & 0.0045 & 51.819 & 0.033 & -0.0443 & 0.0045 & 0.19767 & 0.00033 & 0.33005 & 0.00023 & -4.9868 & 0.0041 \\
2037.71006166 & -25950.23 & 3.65 & 7.2277 & 0.0073 & 51.656 & 0.052 & -0.0425 & 0.0073 & 0.20537 & 0.00070 & 0.33509 & 0.00045 & -4.9608 & 0.0080 \\
2038.71085062 & -25953.11 & 2.80 & 7.2140 & 0.0056 & 51.749 & 0.040 & -0.0527 & 0.0056 & 0.20725 & 0.00049 & 0.33539 & 0.00032 & -4.9597 & 0.0054 \\
2039.71726650 & -25951.96 & 2.22 & 7.2206 & 0.0044 & 51.826 & 0.032 & -0.0418 & 0.0044 & 0.19978 & 0.00033 & 0.32930 & 0.00023 & -4.9829 & 0.0038 \\
2040.70485007 & -25949.89 & 2.78 & 7.2161 & 0.0056 & 51.878 & 0.040 & -0.0562 & 0.0056 & 0.19754 & 0.00039 & 0.32758 & 0.00030 & -4.9741 & 0.0054 \\
2050.69869911 & -25948.66 & 2.38 & 7.2344 & 0.0048 & 51.767 & 0.034 & -0.0456 & 0.0048 & 0.20211 & 0.00049 & 0.33144 & 0.00028 & -4.9300 & 0.0037 \\
2051.69941908 & -25947.75 & 2.28 & 7.2149 & 0.0046 & 51.827 & 0.033 & -0.0531 & 0.0046 & 0.20046 & 0.00032 & 0.32671 & 0.00024 & -4.9405 & 0.0036 \\
2054.72485199 & -25947.00 & 3.27 & 7.2133 & 0.0065 & 51.744 & 0.047 & -0.0440 & 0.0065 & 0.20425 & 0.00058 & 0.32931 & 0.00039 & -4.9329 & 0.0066 \\
2068.61657773 & -25948.94 & 6.50 & 7.2087 & 0.0130 & 51.636 & 0.093 & -0.0475 & 0.0130 & 0.19369 & 0.00076 & 0.31291 & 0.00078 & -5.0547 & 0.0223 \\
2069.66518268 & -25946.55 & 1.78 & 7.2208 & 0.0036 & 51.812 & 0.026 & -0.0387 & 0.0036 & 0.20266 & 0.00030 & 0.33276 & 0.00018 & -4.9176 & 0.0022 \\
2070.70331840 & -25951.22 & 1.90 & 7.2178 & 0.0038 & 51.831 & 0.027 & -0.0422 & 0.0038 & 0.20203 & 0.00030 & 0.33213 & 0.00019 & -4.9771 & 0.0029 \\
2071.70861407 & -25947.49 & 2.27 & 7.2323 & 0.0045 & 51.812 & 0.033 & -0.0404 & 0.0045 & 0.19963 & 0.00032 & 0.32895 & 0.00023 & -4.9530 & 0.0038 \\
2072.70390704 & -25950.16 & 1.92 & 7.2210 & 0.0038 & 51.772 & 0.027 & -0.0519 & 0.0038 & 0.19726 & 0.00023 & 0.32707 & 0.00018 & -4.9343 & 0.0026 \\
2075.64050686 & -25954.65 & 2.07 & 7.2164 & 0.0041 & 51.816 & 0.030 & -0.0435 & 0.0041 & 0.20555 & 0.00039 & 0.33996 & 0.00023 & -4.9387 & 0.0031 \\
2076.71172723 & -25954.10 & 2.86 & 7.2234 & 0.0057 & 51.785 & 0.041 & -0.0508 & 0.0057 & 0.19931 & 0.00049 & 0.33082 & 0.00032 & -4.9489 & 0.0052 \\
2078.68209378 & -25953.69 & 2.11 & 7.2294 & 0.0042 & 51.816 & 0.030 & -0.0537 & 0.0042 & 0.20451 & 0.00042 & 0.33492 & 0.00024 & -4.9276 & 0.0030 \\
2079.69157744 & -25954.09 & 2.05 & 7.2230 & 0.0041 & 51.847 & 0.029 & -0.0500 & 0.0041 & 0.20484 & 0.00033 & 0.33493 & 0.00021 & -4.9673 & 0.0032 \\
2091.68517699 & -25942.99 & 2.22 & 7.2221 & 0.0044 & 51.816 & 0.032 & -0.0457 & 0.0044 & 0.20067 & 0.00033 & 0.33132 & 0.00022 & -4.9330 & 0.0035 \\
2092.62642253 & -25944.84 & 2.75 & 7.2247 & 0.0055 & 51.674 & 0.039 & -0.0514 & 0.0055 & 0.20209 & 0.00043 & 0.33197 & 0.00030 & -4.9541 & 0.0052 \\
2093.65167345 & -25941.22 & 8.21 & 7.1941 & 0.0164 & 51.506 & 0.118 & -0.0339 & 0.0164 & 0.19840 & 0.00154 & 0.33055 & 0.00111 & -4.8667 & 0.0206 \\
2094.60878996 & -25949.64 & 2.19 & 7.2241 & 0.0044 & 51.758 & 0.031 & -0.0438 & 0.0044 & 0.20711 & 0.00052 & 0.34059 & 0.00026 & -4.9491 & 0.0034 \\
2095.61250010 & -25951.13 & 2.51 & 7.2212 & 0.0050 & 51.779 & 0.036 & -0.0429 & 0.0050 & 0.21408 & 0.00084 & 0.34427 & 0.00036 & -4.9968 & 0.0048 \\
2096.63043591 & -25946.18 & 1.86 & 7.2069 & 0.0037 & 51.808 & 0.027 & -0.0468 & 0.0037 & 0.19729 & 0.00025 & 0.32961 & 0.00018 & -4.9417 & 0.0025 \\
2097.66492584 & -25949.58 & 2.68 & 7.2272 & 0.0054 & 51.699 & 0.038 & -0.0420 & 0.0054 & 0.19631 & 0.00038 & 0.32678 & 0.00028 & -4.9684 & 0.0050 \\
2099.64055626 & -25956.66 & 2.95 & 7.2259 & 0.0059 & 51.747 & 0.042 & -0.0436 & 0.0059 & 0.19906 & 0.00050 & 0.33133 & 0.00034 & -4.9475 & 0.0055 \\
2106.65935375 & -25953.42 & 2.74 & 7.2134 & 0.0055 & 51.846 & 0.039 & -0.0509 & 0.0055 & 0.20312 & 0.00046 & 0.33364 & 0.00031 & -4.9325 & 0.0049 \\
2110.71647151 & -25956.36 & 4.06 & 7.2267 & 0.0081 & 51.780 & 0.058 & -0.0364 & 0.0081 & 0.20183 & 0.00080 & 0.33437 & 0.00050 & -5.0276 & 0.0121 \\
2111.55350057 & -25956.94 & 2.17 & 7.2319 & 0.0043 & 51.782 & 0.031 & -0.0471 & 0.0043 & 0.20624 & 0.00072 & 0.34176 & 0.00029 & -4.9499 & 0.0033 \\
2112.58877718 & -25957.19 & 2.32 & 7.2189 & 0.0046 & 51.876 & 0.033 & -0.0499 & 0.0046 & 0.20088 & 0.00053 & 0.33832 & 0.00028 & -4.9460 & 0.0037 \\
2119.69094709 & -25936.50 & 3.00 & 7.2213 & 0.0060 & 51.738 & 0.043 & -0.0435 & 0.0060 & 0.19166 & 0.00032 & 0.31036 & 0.00030 & -4.9406 & 0.0063 \\
2120.67760594 & -25949.18 & 3.00 & 7.2300 & 0.0060 & 51.725 & 0.043 & -0.0433 & 0.0060 & 0.19295 & 0.00036 & 0.31826 & 0.00031 & -4.9643 & 0.0065 \\
2125.54522998 & -25950.83 & 3.03 & 7.2127 & 0.0061 & 51.670 & 0.043 & -0.0465 & 0.0061 & 0.19453 & 0.00040 & 0.32482 & 0.00033 & -4.9352 & 0.0057 \\
2126.56916834 & -25955.91 & 2.91 & 7.2154 & 0.0058 & 51.721 & 0.042 & -0.0540 & 0.0058 & 0.19730 & 0.00047 & 0.32688 & 0.00033 & -4.9425 & 0.0055 \\
2127.63495900 & -25941.46 & 2.34 & 7.1975 & 0.0047 & 51.835 & 0.034 & -0.0335 & 0.0047 & 0.19655 & 0.00029 & 0.32274 & 0.00023 & -4.9525 & 0.0041 \\
2130.59708730 & -25948.12 & 4.13 & 7.2245 & 0.0083 & 51.662 & 0.059 & -0.0473 & 0.0083 & 0.19863 & 0.00069 & 0.32717 & 0.00049 & -4.9716 & 0.0103 \\
2134.62091775 & -25943.35 & 4.14 & 7.2372 & 0.0083 & 51.633 & 0.059 & -0.0463 & 0.0083 & 0.19951 & 0.00062 & 0.32405 & 0.00047 & -4.8843 & 0.0084 \\
2137.55167183 & -25952.16 & 2.29 & 7.2271 & 0.0046 & 51.752 & 0.033 & -0.0377 & 0.0046 & 0.19581 & 0.00036 & 0.33082 & 0.00024 & -4.9229 & 0.0034 \\
2153.56157966 & -25948.65 & 7.52 & 7.2083 & 0.0150 & 50.864 & 0.106 & -0.0698 & 0.0150 & 0.20168 & 0.00131 & 0.33842 & 0.00096 & -5.0351 & 0.0273 \\
2156.54073685 & -25948.62 & 2.32 & 7.2215 & 0.0046 & 51.739 & 0.033 & -0.0405 & 0.0046 & 0.19654 & 0.00027 & 0.32661 & 0.00022 & -4.9489 & 0.0039 \\
2157.58253018 & -25939.99 & 6.69 & 7.2355 & 0.0134 & 51.239 & 0.095 & -0.0508 & 0.0134 & 0.19306 & 0.00113 & 0.33233 & 0.00083 & -4.9111 & 0.0186 \\
2169.31414581 & -25944.61 & 5.37 & 7.2447 & 0.0107 & 51.639 & 0.077 & -0.0544 & 0.0107 & 0.20509 & 0.00097 & 0.35714 & 0.00070 & -5.0666 & 0.0187 \\
2170.34456074 & -25952.45 & 3.49 & 7.2255 & 0.0070 & 51.749 & 0.050 & -0.0483 & 0.0070 & 0.20255 & 0.00051 & 0.33612 & 0.00039 & -4.9288 & 0.0072 \\
2171.31790533 & -25944.67 & 1.99 & 7.2305 & 0.0040 & 51.831 & 0.029 & -0.0426 & 0.0040 & 0.19372 & 0.00023 & 0.32833 & 0.00018 & -4.9499 & 0.0030 \\
2172.31372841 & -25952.72 & 1.73 & 7.2256 & 0.0035 & 51.866 & 0.025 & -0.0493 & 0.0035 & 0.20207 & 0.00025 & 0.33323 & 0.00016 & -4.9514 & 0.0023 \\
2189.39470134 & -25951.29 & 2.60 & 7.2318 & 0.0052 & 51.769 & 0.037 & -0.0409 & 0.0052 & 0.20485 & 0.00039 & 0.33482 & 0.00028 & -4.9447 & 0.0048 \\
2190.35969016 & -25951.01 & 2.46 & 7.2219 & 0.0049 & 51.827 & 0.035 & -0.0447 & 0.0049 & 0.20302 & 0.00041 & 0.34096 & 0.00027 & -4.9638 & 0.0046 \\
2192.33694562 & -25950.74 & 2.22 & 7.2349 & 0.0044 & 51.773 & 0.032 & -0.0394 & 0.0044 & 0.20046 & 0.00035 & 0.33883 & 0.00024 & -4.9831 & 0.0039 \\
2212.37435503 & -25936.78 & 3.64 & 7.2407 & 0.0073 & 51.693 & 0.052 & -0.0481 & 0.0073 & 0.20112 & 0.00050 & 0.33053 & 0.00041 & -4.9130 & 0.0076 \\
2213.42144850 & -25933.17 & 5.10 & 7.2441 & 0.0102 & 51.627 & 0.073 & -0.0424 & 0.0102 & 0.20049 & 0.00081 & 0.34085 & 0.00063 & -4.9514 & 0.0141 \\
2216.40464082 & -25957.30 & 2.52 & 7.2358 & 0.0050 & 51.766 & 0.036 & -0.0469 & 0.0050 & 0.19872 & 0.00033 & 0.33191 & 0.00026 & -4.9473 & 0.0047 \\
2235.36800056 & -25936.90 & 3.27 & 7.2153 & 0.0065 & 51.681 & 0.047 & -0.0392 & 0.0065 & 0.19867 & 0.00039 & 0.32170 & 0.00034 & -4.9430 & 0.0073 \\
2236.31848652 & -25939.04 & 2.01 & 7.2107 & 0.0040 & 51.788 & 0.029 & -0.0386 & 0.0040 & 0.19808 & 0.00022 & 0.32128 & 0.00018 & -4.9093 & 0.0029 \\
2237.32445636 & -25937.89 & 3.21 & 7.2161 & 0.0064 & 51.689 & 0.046 & -0.0480 & 0.0064 & 0.20109 & 0.00045 & 0.33575 & 0.00035 & -4.9572 & 0.0072 \\
2239.31347520 & -25929.04 & 2.48 & 7.2663 & 0.0050 & 51.486 & 0.035 & -0.0354 & 0.0050 & 0.19551 & 0.00027 & 0.33026 & 0.00024 & -4.9109 & 0.0041 \\
2240.31392795 & -25939.68 & 2.49 & 7.2404 & 0.0050 & 51.576 & 0.035 & -0.0372 & 0.0050 & 0.20566 & 0.00039 & 0.34468 & 0.00026 & -4.9716 & 0.0048 \\
2244.31787105 & -25942.97 & 2.97 & 7.2368 & 0.0059 & 51.597 & 0.042 & -0.0408 & 0.0059 & 0.20098 & 0.00049 & 0.33865 & 0.00033 & -4.9580 & 0.0062 \\
2245.32061706 & -25948.66 & 2.68 & 7.2440 & 0.0054 & 51.710 & 0.038 & -0.0410 & 0.0054 & 0.20511 & 0.00042 & 0.33634 & 0.00029 & -4.9453 & 0.0051 \\
2412.68949053 & -25950.92 & 3.05 & 7.2197 & 0.0061 & 51.732 & 0.044 & -0.0454 & 0.0061 & 0.20447 & 0.00062 & 0.33686 & 0.00037 & -4.9377 & 0.0059 \\
2413.66459085 & -25947.71 & 2.65 & 7.2183 & 0.0053 & 51.823 & 0.038 & -0.0441 & 0.0053 & 0.20030 & 0.00040 & 0.32845 & 0.00028 & -4.9284 & 0.0048 \\
2414.68966285 & -25947.48 & 2.49 & 7.2253 & 0.0050 & 51.766 & 0.036 & -0.0377 & 0.0050 & 0.20777 & 0.00070 & 0.33767 & 0.00034 & -4.9543 & 0.0044 \\
2416.63195876 & -25948.72 & 1.84 & 7.2256 & 0.0037 & 51.752 & 0.026 & -0.0450 & 0.0037 & 0.20193 & 0.00025 & 0.33440 & 0.00017 & -4.9112 & 0.0024 \\
2417.65217737 & -25945.45 & 2.23 & 7.2299 & 0.0045 & 51.679 & 0.032 & -0.0452 & 0.0045 & 0.20080 & 0.00033 & 0.33441 & 0.00023 & -4.9182 & 0.0034 \\
2418.65567006 & -25939.90 & 4.63 & 7.2226 & 0.0093 & 51.583 & 0.066 & -0.0491 & 0.0093 & 0.20035 & 0.00082 & 0.33348 & 0.00058 & -4.8454 & 0.0094 \\
2427.73262001 & -25948.08 & 2.51 & 7.2244 & 0.0050 & 51.677 & 0.036 & -0.0372 & 0.0050 & 0.20136 & 0.00034 & 0.32694 & 0.00025 & -4.9256 & 0.0042 \\
2428.66099085 & -25948.20 & 3.48 & 7.2369 & 0.0070 & 51.680 & 0.050 & -0.0446 & 0.0070 & 0.19891 & 0.00059 & 0.33490 & 0.00040 & -4.9245 & 0.0073 \\
2430.67069466 & -25949.69 & 2.69 & 7.2243 & 0.0054 & 51.826 & 0.039 & -0.0456 & 0.0054 & 0.20210 & 0.00049 & 0.33403 & 0.00030 & -4.9424 & 0.0051 \\
2431.71441182 & -25953.31 & 3.01 & 7.2281 & 0.0060 & 51.773 & 0.043 & -0.0573 & 0.0060 & 0.20348 & 0.00051 & 0.33419 & 0.00034 & -4.9562 & 0.0063 \\
2443.66167945 & -25939.53 & 4.19 & 7.2331 & 0.0084 & 51.677 & 0.060 & -0.0390 & 0.0084 & 0.19808 & 0.00094 & 0.33865 & 0.00055 & -4.9262 & 0.0096 \\
2444.58062692 & -25940.18 & 3.73 & 7.2304 & 0.0075 & 51.632 & 0.053 & -0.0418 & 0.0075 & 0.20146 & 0.00072 & 0.33526 & 0.00046 & -4.9054 & 0.0077 \\
2445.59714044 & -25938.62 & 2.63 & 7.2310 & 0.0053 & 51.704 & 0.038 & -0.0521 & 0.0053 & 0.20681 & 0.00060 & 0.33816 & 0.00032 & -4.9227 & 0.0045 \\
2446.60954406 & -25943.91 & 2.04 & 7.2346 & 0.0041 & 51.707 & 0.029 & -0.0401 & 0.0041 & 0.21010 & 0.00047 & 0.34039 & 0.00023 & -4.9044 & 0.0028 \\
2447.61816721 & -25946.79 & 2.45 & 7.2228 & 0.0049 & 51.731 & 0.035 & -0.0402 & 0.0049 & 0.20634 & 0.00047 & 0.33573 & 0.00028 & -4.9201 & 0.0040 \\
2448.59247450 & -25943.51 & 1.99 & 7.2276 & 0.0040 & 51.736 & 0.028 & -0.0471 & 0.0040 & 0.20433 & 0.00039 & 0.33905 & 0.00021 & -4.9295 & 0.0028 \\
2449.61747375 & -25952.32 & 2.00 & 7.2316 & 0.0040 & 51.709 & 0.029 & -0.0501 & 0.0040 & 0.20561 & 0.00039 & 0.33836 & 0.00022 & -4.9469 & 0.0030 \\
2453.58219427 & -25954.60 & 2.51 & 7.2317 & 0.0050 & 51.731 & 0.036 & -0.0508 & 0.0050 & 0.20724 & 0.00054 & 0.33398 & 0.00030 & -4.9429 & 0.0044 \\
2454.66812667 & -25947.21 & 6.43 & 7.1978 & 0.0129 & 51.332 & 0.092 & -0.0655 & 0.0129 & 0.20930 & 0.00134 & 0.32552 & 0.00083 & -4.8716 & 0.0148 \\
2455.68216977 & -25945.60 & 2.35 & 7.2287 & 0.0047 & 51.761 & 0.034 & -0.0504 & 0.0047 & 0.19603 & 0.00032 & 0.32519 & 0.00023 & -4.9350 & 0.0040 \\
2456.70439482 & -25947.19 & 2.27 & 7.2287 & 0.0045 & 51.760 & 0.032 & -0.0431 & 0.0045 & 0.20024 & 0.00032 & 0.32957 & 0.00023 & -4.9428 & 0.0037 \\
2457.65525719 & -25948.19 & 1.99 & 7.2315 & 0.0040 & 51.728 & 0.028 & -0.0453 & 0.0040 & 0.20246 & 0.00029 & 0.33203 & 0.00020 & -4.9084 & 0.0027 \\
2458.62496853 & -25945.07 & 2.27 & 7.2291 & 0.0045 & 51.777 & 0.033 & -0.0417 & 0.0045 & 0.20554 & 0.00043 & 0.33599 & 0.00025 & -4.9456 & 0.0037 \\
2459.58949268 & -25951.90 & 2.95 & 7.2230 & 0.0059 & 51.780 & 0.042 & -0.0467 & 0.0059 & 0.20732 & 0.00081 & 0.33773 & 0.00040 & -4.9534 & 0.0058 \\
2460.68089277 & -25942.78 & 2.17 & 7.2312 & 0.0043 & 51.747 & 0.031 & -0.0484 & 0.0043 & 0.19974 & 0.00027 & 0.32386 & 0.00021 & -4.9177 & 0.0033 \\
2461.61328194 & -25950.89 & 2.38 & 7.2216 & 0.0048 & 51.758 & 0.034 & -0.0482 & 0.0048 & 0.20643 & 0.00042 & 0.33320 & 0.00026 & -4.9397 & 0.0040 \\
2462.59779913 & -25952.54 & 2.76 & 7.2323 & 0.0055 & 51.768 & 0.040 & -0.0544 & 0.0055 & 0.20247 & 0.00044 & 0.33246 & 0.00030 & -4.9271 & 0.0050 \\
2464.59168365 & -25943.37 & 6.05 & 7.2300 & 0.0121 & 51.681 & 0.087 & -0.0550 & 0.0121 & 0.20081 & 0.00129 & 0.33041 & 0.00082 & -5.0067 & 0.0196 \\
2465.55365903 & -25950.80 & 2.57 & 7.2260 & 0.0051 & 51.679 & 0.037 & -0.0511 & 0.0051 & 0.20878 & 0.00047 & 0.33229 & 0.00028 & -4.9127 & 0.0043 \\
2472.64517694 & -25941.76 & 3.16 & 7.2359 & 0.0063 & 51.753 & 0.045 & -0.0520 & 0.0063 & 0.19636 & 0.00045 & 0.33049 & 0.00034 & -4.9418 & 0.0068 \\
2473.58292067 & -25943.18 & 4.22 & 7.2244 & 0.0084 & 51.761 & 0.061 & -0.0520 & 0.0084 & 0.20374 & 0.00082 & 0.33807 & 0.00053 & -5.0288 & 0.0123 \\
2475.57164684 & -25941.40 & 2.96 & 7.2175 & 0.0059 & 51.614 & 0.042 & -0.0479 & 0.0059 & 0.20318 & 0.00046 & 0.33367 & 0.00033 & -4.8946 & 0.0055 \\
2476.56363715 & -25940.68 & 2.08 & 7.2369 & 0.0042 & 51.694 & 0.030 & -0.0435 & 0.0042 & 0.20135 & 0.00031 & 0.33435 & 0.00020 & -4.9312 & 0.0032 \\
2477.54328633 & -25943.82 & 2.36 & 7.2267 & 0.0047 & 51.619 & 0.034 & -0.0477 & 0.0047 & 0.20568 & 0.00040 & 0.33665 & 0.00025 & -4.9244 & 0.0039 \\
2478.54019115 & -25942.74 & 2.84 & 7.2352 & 0.0057 & 51.520 & 0.040 & -0.0336 & 0.0057 & 0.20155 & 0.00045 & 0.33490 & 0.00031 & -4.8846 & 0.0049 \\
2479.53140970 & -25947.93 & 2.57 & 7.2379 & 0.0051 & 51.599 & 0.037 & -0.0496 & 0.0051 & 0.20166 & 0.00038 & 0.33160 & 0.00026 & -4.9276 & 0.0046 \\
2481.49463114 & -25946.30 & 3.03 & 7.2289 & 0.0061 & 51.619 & 0.043 & -0.0492 & 0.0061 & 0.20294 & 0.00048 & 0.33199 & 0.00033 & -4.9290 & 0.0060 \\
2513.45063116 & -25941.24 & 2.40 & 7.2144 & 0.0048 & 51.895 & 0.035 & -0.0373 & 0.0048 & 0.20640 & 0.00044 & 0.33944 & 0.00025 & -4.9362 & 0.0047 \\
2513.50299346 & -25947.17 & 3.03 & 7.2025 & 0.0061 & 51.855 & 0.044 & -0.0382 & 0.0061 & 0.20459 & 0.00053 & 0.33408 & 0.00032 & -4.9517 & 0.0073 \\
2515.41998002 & -25948.56 & 2.34 & 7.2146 & 0.0047 & 51.939 & 0.034 & -0.0431 & 0.0047 & 0.20636 & 0.00039 & 0.33775 & 0.00024 & -4.9439 & 0.0045 \\
2515.44534089 & -25950.25 & 1.94 & 7.2047 & 0.0039 & 51.950 & 0.028 & -0.0464 & 0.0039 & 0.20134 & 0.00030 & 0.34026 & 0.00018 & -4.9263 & 0.0032 \\
2516.53158394 & -25947.31 & 3.15 & 7.2198 & 0.0063 & 51.853 & 0.045 & -0.0425 & 0.0063 & 0.20164 & 0.00051 & 0.33378 & 0.00033 & -4.9089 & 0.0073 \\
2565.36730046 & -25946.28 & 1.77 & 7.2193 & 0.0035 & 51.900 & 0.025 & -0.0456 & 0.0035 & 0.20431 & 0.00030 & 0.33899 & 0.00017 & -4.9267 & 0.0028 \\
2566.34623047 & -25947.65 & 2.17 & 7.2162 & 0.0043 & 51.897 & 0.031 & -0.0510 & 0.0043 & 0.20611 & 0.00038 & 0.33952 & 0.00022 & -4.9266 & 0.0039 \\
2575.40577181 & -25940.30 & 2.32 & 7.1948 & 0.0046 & 51.938 & 0.034 & -0.0472 & 0.0046 & 0.19944 & 0.00032 & 0.32769 & 0.00022 & -4.9057 & 0.0044 \\
2579.38604962 & -25941.06 & 2.55 & 7.2168 & 0.0051 & 51.803 & 0.037 & -0.0471 & 0.0051 & 0.20663 & 0.00039 & 0.33812 & 0.00025 & -4.9279 & 0.0054 \\
2580.41346947 & -25943.54 & 5.80 & 7.2096 & 0.0116 & 51.886 & 0.084 & -0.0440 & 0.0116 & 0.21583 & 0.00128 & 0.33656 & 0.00073 & -5.0510 & 0.0261 \\
2584.34061165 & -25943.38 & 3.56 & 7.2126 & 0.0071 & 51.855 & 0.051 & -0.0389 & 0.0071 & 0.20706 & 0.00065 & 0.34449 & 0.00040 & -4.9638 & 0.0099 \\
2588.34564266 & -25941.61 & 5.73 & 7.2024 & 0.0115 & 51.732 & 0.082 & -0.0406 & 0.0115 & 0.21510 & 0.00104 & 0.33994 & 0.00069 & -5.1394 & 0.0308 \\
2601.31755833 & -25939.58 & 2.08 & 7.2186 & 0.0042 & 51.878 & 0.030 & -0.0441 & 0.0042 & 0.21007 & 0.00031 & 0.33386 & 0.00020 & -4.8917 & 0.0033 \\
3574.66373124 & -25937.49 & 2.73 & 7.2156 & 0.0055 & 51.803 & 0.039 & -0.0596 & 0.0055 & 0.20092 & 0.00041 & 0.33568 & 0.00028 & -4.9432 & 0.0058 \\
3576.67067116 & -25940.31 & 4.33 & 7.1878 & 0.0087 & 52.058 & 0.063 & -0.0508 & 0.0087 & 0.20641 & 0.00067 & 0.33027 & 0.00048 & -4.9307 & 0.0123 \\
3584.66117572 & -25952.06 & 3.29 & 7.2038 & 0.0066 & 51.971 & 0.047 & -0.0413 & 0.0066 & 0.19382 & 0.00050 & 0.33094 & 0.00034 & -4.9184 & 0.0078 \\
3593.35139703 & -25946.50 & 2.43 & 7.2084 & 0.0049 & 51.952 & 0.035 & -0.0482 & 0.0049 & 0.19950 & 0.00039 & 0.33175 & 0.00024 & -5.0023 & 0.0057 \\
3594.39633670 & -25951.08 & 1.96 & 7.2063 & 0.0039 & 51.962 & 0.028 & -0.0478 & 0.0039 & 0.20373 & 0.00030 & 0.33283 & 0.00018 & -4.9150 & 0.0031 \\
3619.43182686 & -25954.25 & 2.15 & 7.2178 & 0.0043 & 51.955 & 0.031 & -0.0551 & 0.0043 & 0.20830 & 0.00051 & 0.34274 & 0.00024 & -4.9483 & 0.0039 \\
3626.43584414 & -25955.03 & 2.33 & 7.2124 & 0.0047 & 51.913 & 0.034 & -0.0462 & 0.0047 & 0.20240 & 0.00043 & 0.34080 & 0.00024 & -4.9182 & 0.0041 \\
3628.45057967 & -25947.93 & 4.67 & 7.1976 & 0.0093 & 51.703 & 0.067 & -0.0418 & 0.0093 & 0.20078 & 0.00091 & 0.35196 & 0.00057 & -4.9337 & 0.0133 \\
3652.37335477 & -25940.45 & 2.96 & 7.2105 & 0.0059 & 51.918 & 0.043 & -0.0457 & 0.0059 & 0.20705 & 0.00066 & 0.33466 & 0.00036 & -4.9787 & 0.0067 \rule[-0.8ex]{0pt}{0pt} \\

    \bottomrule
\end{tabular}}}
\end{table*}

\begin{table}
\centering
\caption{Prior parameter distribution for the calculation of the interior compositions of TOI-1422\,c and TOI-1422\,b where we solve for atmospheric mass, the mass of the rocky interior (comprising core and mantle), and $Z_\mathrm{env}$ the envelope metallicity.}\label{tab:comppriors}
\renewcommand{\arraystretch}{1.2}
\begin{tabular}{lcc}
    \hline\hline
     Parameter & Prior range & Distribution \\
    \hline \\[-6pt]%
    $M_\mathrm{atm}$\dotfill & $(0.01 - 0.15)*M_{p}$ & log-uniform \\
    $M_\mathrm{core+mantle}$\dotfill & $\mathcal{N}(M_p, \sigma^2_{M_p})$  & gaussian  \\
    $Z$\dotfill & $0.02-1.0$ & uniform\\
    \bottomrule
\end{tabular}
\end{table}

\bsp	
\label{lastpage}
\end{document}